\documentclass[twoside,11pt]{article}

\usepackage{jmlr2e}

\usepackage{microtype}
\usepackage{graphicx}
\usepackage{subfigure}
\usepackage{booktabs} %

\usepackage{hyperref}

\usepackage{placeins} %
\usepackage{amsmath}
\allowdisplaybreaks

\jmlrheading{~}{~}{~}{13th July 2021}{~}{~}{~}

\ShortHeadings{Kernel-Matrix Determinant Estimates from stopped Cholesky Decomposition}{Bartels, Boomsma, Frellsen and Garreau}
\firstpageno{1}

\usepackage{microtype,marvosym} 
\usepackage{stmaryrd} 
\usepackage{amsmath,amssymb,graphicx}
\usepackage{url}

\usepackage{afterpage}
\usepackage{nameref}
\usepackage{adjustbox}

\usepackage{pgfplots}
\pgfplotsset{compat=newest}

\pgfplotsset{
	every axis legend/.append style =
	{
		cells = { anchor = east },
		draw  = none
	},
}  
\makeatletter
\pgfplotsset{ 
	range frame/.style={
		tick align = outside,
		axis line style={opacity=0},
		after end axis/.code={
			\draw ({rel axis cs:0,0}-|{axis cs:\pgfplots@data@xmin,0}) -- ({rel axis cs:0,0}-|{axis cs:\pgfplots@data@xmax,0});
			\draw ({rel axis cs:0,0}|-{axis cs:0,\pgfplots@data@ymin}) -- ({rel axis cs:0,0}|-{axis cs:0,\pgfplots@data@ymax});
		}
	}
}
\makeatother

\pgfkeys{/pgfplots/mytuftestyle/.style={
		semithick,
		tick style={major tick length=4pt,semithick,black},
		separate axis lines,
		axis x line*=bottom,
		axis x line shift=5pt,
		xlabel shift=0pt,
		axis y line*=left,
		tick align = outside,
		axis y line shift=5pt,
		ylabel shift=0pt}}

\usepgfplotslibrary{external} 
\tikzsetexternalprefix{tikz_out/}
\usetikzlibrary{plotmarks}
\usepgfplotslibrary{groupplots}
\pgfplotsset{plot coordinates/math parser=false}       
\pgfkeys{/pgfplots/mystyle/.style={
		semithick,
		tick style={major tick length=4pt,semithick,gray},
		xtick align = inside,
		ytick align = inside,
		xlabel near ticks,
		ylabel near ticks,
}}
\pgfkeys{/pgfplots/resultfigstyle/.style={
		semithick,
		tick style={major tick length=4pt,semithick,gray},
		xtick align = inside,
		xticklabel style = {align=center},
		ytick align = inside,
		xlabel near ticks,
		ylabel near ticks,
		every axis x label/.style={
				at={(ticklabel cs:-0.05, -20)},anchor=near ticklabel
			},
		every axis y label/.style={
				at={(ticklabel cs:1, -25)},anchor=near ticklabel
		},
		every tick label/.append style={font=\tiny}		
}}

\newlength\figureheight
\newlength\figurewidth
\newlength\figheight
\newlength\figwidth

\usepackage{natbib}
\newcommand{\textcite}{\citet}
\newcommand{\parencite}{\citep}

\definecolor{lred}{RGB}{200,0,0}
\definecolor{dred}{RGB}{130,0,0} \definecolor{dblu}{RGB}{0,0,130}
\definecolor{dgre}{RGB}{0,130,0} \definecolor{dgra}{RGB}{50,50,50}
\definecolor{mgra}{RGB}{100,100,100}
\definecolor{lgra}{RGB}{220,220,220}
\definecolor{MPG}{RGB}{000,125,122}
\definecolor{mpg}{RGB}{000,125,122}
\definecolor{ora}{HTML}{FF9933}
\definecolor{newcode}{RGB}{220,220,220}

\definecolor{AMPurple}{HTML}{663366}
\definecolor{Burgundy}{HTML}{993333}
\definecolor{Coffee}{HTML}{7B6049}
\definecolor{ForestGreen}{HTML}{005826}
\definecolor{Lavender}{HTML}{6E6AB1}
\definecolor{PSLightBlue}{HTML}{7DA7D9}

\usepackage{booktabs}

\newcommand{\Exp}{\mathbb{E}}
\newcommand{\Var}{\mathbb{V}}

\renewcommand{\Re}{\mathbb{R}}

\newcommand{\one}{\mathbf{1}}

\newcommand{\N}{\mathcal{N}}

\renewcommand{\L}{\mathcal{L}}
\newcommand{\Trans}{^{\intercal}} 
\newcommand{\iTrans}{^{-\intercal}}

\renewcommand{\det}{\operatorname{det}}

\renewcommand{\=}{\operatorname*{=}}

\renewcommand{\vec}{\boldsymbol} 
\newcommand{\mat}{\boldsymbol} 
\newcommand{\inv}[1]{{#1}^{-\!1}}
\newcommand{\trace}{\operatorname{trace}}

\renewcommand{\O}{\mathcal{O}}

\renewcommand{\L}{\mathcal{L}}

\newcommand{\X}{\mathbb{X}}

\usepackage{colonequals}
\newcommand{\ce}{\colonequals}

\usepackage{nicefrac}

\usetikzlibrary{arrows,shapes,plotmarks}

\tikzset{>=stealth'} 
\tikzstyle{graphnode} = 
[circle,draw=black,minimum size=22pt,text centered,text
width=22pt,inner sep=0pt] 
\tikzstyle{var}   =[graphnode,fill=white]
\tikzstyle{obs}   =[graphnode,fill=black,text=white]
\tikzstyle{fac}   =[rectangle,draw=black,fill=black!25,minimum size=5pt]
\tikzstyle{facprior} =[rectangle,draw=black,fill=black,text=white,minimum size=5pt]
\tikzstyle{edge}  =[draw=white,double=black,thick,-]
\tikzstyle{prior} =[rectangle, draw=black, fill=black, minimum size=
5pt, inner sep=0pt]
\tikzstyle{dirprior} = [circle, draw=black, fill=black, minimum
size=5pt, inner sep=0pt]

\DeclareSymbolFont{stmry}{U}{stmry}{m}{n}
\DeclareMathSymbol\leftarrowtriangle\mathrel{stmry}{"5E}
\DeclareMathSymbol\rightarrowtriangle\mathrel{stmry}{"5F}
\renewcommand{\gets}{\operatorname*{\leftarrowtriangle}}

\newcounter{PHcomment}

\usepackage{algorithm}
\usepackage{algpseudocode}

\makeatletter
\newcommand{\superimpose}[2]{
	{\ooalign{$#1\@firstoftwo#2$\cr\hfil$#1\@secondoftwo#2$\hfil\cr}}}
\makeatother

\usepackage{epigraph}
\usepackage{amsfonts}
\usepackage{mdframed}
\renewcommand{\epsilon}{\varepsilon}

\newcommand{\eqcomment}[1]{\\*&\qquad(\text{{\small\emph{#1}}})\notag}

\newcommand{\norm}[1]{\|#1\|}

\newcommand{\pname}[1]{\emph{#1}} 
\newcommand{\latin}[1]{\emph{#1}} 

\newcommand{\eg}{\latin{e.g.~}}

\renewcommand{\Re}{\mathbb{R}}
\newcommand{\reals}{\Re}
\newcommand{\sign}{\operatorname{sign}}

\renewcommand{\epsilon}{\varepsilon}

\newcommand{\abs}[1]{\left\lvert#1\right\rvert} 
\newcommand{\smallabs}[1]{\lvert#1\rvert}
\renewcommand{\det}[1]{\mathrm{det}\left(#1\right)}
\newcommand{\pn}[1]{}
\newcommand{\Proba}{\mathbb{P}}
\newcommand{\proba}[1]{\Proba\left (#1\right )}
\newcommand{\marginnote}[1]{}

\newcommand{\dx}[1]{\mathrm{d}#1}

\renewcommand{\l}[1]{f_{#1}} %
\newcommand{\lBound}{C} 
\newcommand{\me}[1][\tau]{\hat{\mu}_{#1}} 
\newcommand{\lDetn}[1]{D_{#1}} 
\newcommand{\kj}[1][j]{\vec k_{#1}(\vec x)} 
\renewcommand{\L}{\mathcal{L}}
\newcommand{\U}{\mathcal{U}}
\newcommand{\stopTimeOne}{\sign(\U_n)=\sign(\L_n)\neq 0}
\newcommand{\stopTimeTwo}{\frac{\U_n-\L_n}{2\min(\abs{\U_n}, \abs{\L_n})} \leq r}

\makeatletter
\def\th@plain{%
	\thm@notefont{}
	\itshape 
}
\def\th@definition{%
	\thm@notefont{}
	\normalfont 
}
\makeatother

\newcommand{\metric}{m}

\newcommand{\expconfig}[3]{#3}
\newcommand{\labely}{$\nicefrac{\U_n-\L_n}{2\min(\abs{\U_n}, \abs{\L_n})}$}
\newcommand{\labelx}{~}

\newcommand{\ticklabelsx}{true}
\newcommand{\ticklabelsy}{true}
\newcommand{\METRO}{METRO}
\newcommand{\PM}{PM2.5}
\newcommand{\TAMILNADU}{TAMILNADU}
\newcommand{\PROTEIN}{PROTEIN}
\newcommand{\BANK}{BANK}
\newcommand{\PUMADYN}{PUMADYN}
\newcommand{\overhead}{$5\%$ overhead}
\newcommand{\DefaultChol}{default$\quad$}
\newcommand{\StoppedChol}{probabilistic$\quad$}
\newcommand{\PivotedChol}{pivoting$\quad$}
\newcommand{\DiagonalPrecision}{diagonal precision$\quad$}
\newcommand{\meantime}[1]{\colorbox{white}{\textbf{\large mean time default Cholesky: #1 s}}}
\newcommand{\clearlegend}{}

\newcommand{\hidelegend}{\renewcommand{\clearlegend}{\legend{}}}
\newcommand{\runinfo}[7]{$\ell=#6$}

\newcommand{\augmentedStatement}[1]{\colorbox{newcode}{{#1}}}
\newcommand{\AState}[1]{\State \augmentedStatement{#1}} 
\newcommand{\indk}{j}
\newcommand{\indj}{k}
\newcommand{\indi}{i}

\newcommand{\captionText}[1]{
	relative execution times to compute the log-determinant using RBF (\textbf{left panel}) and OU (\textbf{right panel}) kernels on the #1 dataset for $\theta=1$, $\log\ell=-1, \dots, 3$ and $\delta=0.1$ for ten repetitions.
	The number next to one on the $y$-axis displays the absolute execution times of the default Cholesky.
	The solid, horizontal, orange line (\ref{leg:overhead}) visualizes the $105\%$ mark.
	The $x$-axis displays a desired absolute precision on the diagonal elements $d$ (top) and the average corresponding desired relative precision~$r$ (bottom) on the log-determinant.
}%

\usepackage[capitalize]{cleveref} 
\usepackage{autonum} 

\usepackage{xr}  %

\newcommand{\STATE}{\State}

\newcommand{\IF}{\If}
\newcommand{\ENDIF}{\EndIf}
\newcommand{\COMMENT}[1]{#1}

\begin{document}
\title{Kernel-Matrix Determinant Estimates from stopped Cholesky Decomposition}

\author{\name Simon Bartels \email bartels@di.ku.dk\\
	\addr
	University of Copenhagen\\
	Universitetsparken 1\\
	2100 K\o{}benhavn, Denmark
	\AND
	\name Wouter Boomsma \email wb@di.ku.dk \\
	\addr
	University of Copenhagen\\
	Universitetsparken 1\\
	2100 K\o{}benhavn, Denmark
	\AND
	\name Jes Frellsen \email jefr@dtu.dk \\
	\addr 
	Technical University of Denmark\\
	Richard Petersens Plads\\
	2800 Kgs.~Lyngby, Denmark
	\AND
	\name Damien Garreau \email damien.garreau@unice.fr\\
	\addr
	Universit\'e C\^ote d'Azur, Inria, CNRS, LJAD\\
	Parc Valrose\\
	06108 Nice Cedex 2, France}

\editor{~}

\maketitle

\begin{abstract}
	Algorithms involving Gaussian processes or determinantal point processes typically require computing the determinant of a kernel matrix.
	Frequently, the latter is computed from the Cholesky decomposition, an algorithm of cubic complexity in the size of the matrix. 
	We show that, under mild assumptions, it is possible to estimate the determinant from only a sub-matrix, with probabilistic guarantee on the relative error. 
	We present an augmentation of the Cholesky decomposition that stops under certain conditions before processing the whole matrix.
	Experiments demonstrate that this can save a considerable amount of time while having an overhead of less than $5\%$ when not stopping early.
	More generally, we present a probabilistic stopping strategy for the approximation of a sum of known length where addends are revealed sequentially. 
	We do not assume independence between addends, only that they are bounded from below and decrease in conditional expectation. 
\end{abstract}
\begin{keywords}
	Gaussian Processes, Optimal Stopping, Kernel Methods, Kriging
\end{keywords}
\tikzset{external/force remake=false}%
\setlength{\figwidth}{.5\textwidth}%
\setlength{\figheight}{.25\textheight}%
\tikzexternaldisable%
{\hspace{-30cm}\smash{\input{tikz/legend.tikz}}}%
{\hspace{-30cm}\smash{\input{tikz/bound_legend.tikz}}}%
\tikzexternalenable%
\definecolor{color1}{rgb}{1,0.6,0.2}%
\definecolor{color0}{rgb}{0,0.4717,0.4604}%
\newcommand{\resultfigheight}{.16\textheight}%
\newcommand{\resultfigwidth}{.5\textwidth}%
{\hspace{-30cm}\smash{\cref{eq:det_estimator}}%
\section{Introduction}

Gaussian processes are a popular probabilistic model in the machine learning community, and a core element of many other methods such as Bayesian optimization \parencite{Mockus1975BO}, Bayesian quadrature \parencite{diaconis88:_bayes}, probabilistic numerics \citep{HenOsbGirRSPA2015} or the \pname{Automatic Statistician} \citep{Steinruecken2019automaticStatistician}.
Typically, inference with a Gaussian process requires the computation of a Cholesky decomposition of a kernel matrix.
For most datasets, this is computationally feasible despite the cubic worst-case complexity of the Cholesky decomposition in the number of samples. 
Nevertheless, when this computation has to performed often, \emph{e.g.}, to optimize kernel parameters, the computational cost of this decomposition becomes paramount. 
When a kernel's parameters do not fit well with the data, our observation is that the log-determinant of the kernel matrix can often be predicted from a subset.
This situation frequently occurs in particular at the beginning of the kernel-parameter optimization process.
In the following, we will demonstrate that it is possible (i) to recognize this situation while computing the Cholesky decomposition, and (ii) to stop the computation prematurely, which can save a considerable amount of time. 
When we are not in a situation that justifies stopping the computation early, we propose to simply continue the computation of the log-determinant until the end. 
Thus the additional computational cost of our method is just that of keeping track of some simple numerical indicators. 
The main benefit of our method is that it provides an ``almost-free lunch'' since \textbf{the overhead when not stopping early is relatively small} (on average less than five percent).
To make this idea practical, we modified the OpenBLAS \citep{Wang2013OpenBLAS} implementation and made our code\footnote{\url{https://github.com/SimonBartels/pac_kernel_matrix_determinant_estimation}} available.

More generally, we will see that our optional stopping strategy can be used to estimate a sum of random variables that are decreasing in expectation. 
In this general setting, we prove that our stopped Cholesky decomposition returns an estimate of a desired relative precision~$r$ with respect to the full computation, with probability $1-\delta$, where~$\delta$ is a user-defined probability threshold. 
For a given level of accuracy that is satisfactory for the problem at hand, the user can then pick a level of confidence in the result and obtain a gain in computational cost with provable guarantees. 
The level of confidence is \textbf{the only parameter} of our method.

\begin{figure}
	{\centering\ \ \ref{leg:bound}\\}%
	\setlength{\figwidth}{.5\textwidth}%
	\setlength{\figheight}{\figwidth}%
	\setlength{\tabcolsep}{-3pt}%
	\begin{tabular}[t]{lr}
		\renewcommand{\labelx}{processed datapoints}%
		\renewcommand{\labely}{relative error bound}%
		\renewcommand{\ticklabelsy}{true}%
		\hidelegend{}%
		\renewcommand{\runinfo}[7]{\textbf{#1} (D=#2)}%
		\adjustbox{valign=t}{\input{tikz/RBF/tamilnadu_electricity/bound_-1.0.tikz}}%
		&
		\renewcommand{\runinfo}[7]{}%
		\renewcommand{\labelx}{$\frac{t}{\operatorname{mean}(t_{\text{default}})}$}%
		\renewcommand{\labely}{~}%
		\adjustbox{valign=t}{\input{tikz/RBF/tamilnadu_electricity/inverse_plot_-1.0.tikz}}%
	\end{tabular}
	\caption{an early stopping scenario.
		We compute the log-determinant of a kernel matrix using the RBF kernel (with $\theta=1$, $\ell=\exp(-1)$ in Eq.~\eqref{eq:kernel_rbf}) on the \TAMILNADU{} dataset (see \cref{tbl:datasets}) for ten random permutations of the dataset.
		\\
		\textbf{Left panel:} 
		progression of our stopping condition (\cref{eq:condition_relative_one} with $\delta\ce 0.1$), as solid, green lines.
		The variance between repetitions is so small such that only one line is visible to the eye.
		We compare against approximate Cholesky decomposition with pivoting (\ref{leg:PivotedChol}, see \cref{sec:baseline}) and mark its stopping points with red crosses.
		The horizontal lines (\ref{leg:DiagonalPrecision}) mark the mean relative precision corresponding to an absolute approximation error on the diagonal elements (denoted with $d$) which is the pivoted Cholesky's stopping criterion.
		\\
		Even for such a short length-scale $\ell$ and a desired relative error $r=0.1$, our algorithm touches only half the dataset before stopping.
		The singularity in the beginning of the stopping condition stems from the denominator crossing $0$ which demonstrates the necessity of the second stopping condition \cref{eq:condition_sign_one}.
		The reason for the slope changes are switches from the deterministic bound $\U''_n$ to $\U'_n$ and back in \cref{eq:bound_final_form}.
		\\
		\textbf{Right panel:} for each repetition, fraction of both algorithm's CPU time over the mean time of the default Cholesky.
		Since $S$ steps of the approximate Cholesky with pivoting cost $\O(NS^2)$ operations, it stops earlier in the left panel, but our algorithm (\ref{leg:StoppedChol}) scaling as $\O(S^3)$ is faster in practice.
	}
	\label{fig:results_bound_early_stop}
\end{figure}

\section{Problem Setup, Related Work and Background}
\label{sec:cholesky}

\subsection{Problem setup}

Given a $\sigma^2\in\Re^+$, a set of inputs $\vec x_1, \ldots, \vec x_N\in \X$ and a kernel function $k : \X\times\X\rightarrow\reals$, we define the \emph{kernel matrix} $\mat A\ce \mat K_N+\sigma^2\mat I_N$, where
\begin{align}
&&\mat K_N\ce\begin{bmatrix}
k(\vec x_1, \vec x_1) & k(\vec x_1, \vec x_2) & \dots & k(\vec x_1, \vec x_N)\\
k(\vec x_2, \vec x_1) & k(\vec x_2, \vec x_2) & & \vdots \\
\vdots & & \ddots & \\
k(\vec x_N, \vec x_1) & \dots  & & k(\vec x_N, \vec x_N)
\end{bmatrix}
\, .
\end{align}
The main focus of this article is the efficient computation of $
\log \det{\mat A}$, which is typically achieved via Cholesky decomposition of~$\mat A$, if $N$ is not too large.
That is, find the unique, lower triangular matrix $\mat C\in\reals^{N\times N}$ satisfying $\mat C\mat C\Trans = \mat A$.
Given the Cholesky decomposition of~$\mat A$, one subsequently computes the log-determinant using the formula
\[
\log \det{\vec A}=2\sum_{n=1}^N \log \mat C_{nn}
\, .
\]
\subsection{Related work}
\label{sec:related_work}
Approximation methods for the log-determinant have been studied extensively---often the more general case of symmetric and positive definite matrices \citep{Skilling1989Eigenvalues,Seeger2000Skilling, Dorn2015Determinant,Ubaru2017fastTrA,Fitzsimons2017BayesianDeterminant,Fitzsimons2017EntropicDeterminant,Saibaba2017Determinant,Boutsidis2017Determinant,Dong2017scalable,Gardner2018gpytorch}.
All of the aforementioned methods are conceptually similar in that they rely on stochastic trace estimators:
the kernel matrix is multiplied with random (probe) vectors and the inner products of the results are used to construct an estimate for the log-determinant.
The theoretical performance analysis of these methods often requires knowledge or an upper bound on expensive-to-compute quantities such as the largest eigenvalue, the condition number or eigenvalue gaps of~$\mat A$  \citep{Ubaru2017fastTrA,Boutsidis2017Determinant,Saibaba2017Determinant,Gardner2018gpytorch}. An advantage of our approach is that we only require knowledge of the largest diagonal entry on~$\mat A$ and a lower bound on the smallest eigenvalue which is given by~$\sigma^2$.

Most related to our work are \citet{Ubaru2017fastTrA,Boutsidis2017Determinant,Gardner2018gpytorch} in the sense that for a desired relative precision and confidence, they proof how to set the parameters of their algorithms accordingly.
Though, a noteworthy distinction to our work is the choice of the probability measure which the desired confidence refers to.
In our case, this probability measure is the law of the inputs~$\vec x_i$.
For the stochastic trace estimators the confidence refers to the source of randomness of the probe vectors.
For the problems we consider in our experiments in \cref{sec:experiments}, none of the theorems by \citet{Boutsidis2017Determinant,Ubaru2017fastTrA,Gardner2018gpytorch} that guarantee relative precision are applicable.
Lemma~8 by \citet{Boutsidis2017Determinant} assumes that all eigenvalues are bounded from above by~1.
This assumption can be established by dividing~$\mat A$ by~$\trace[\mat A]$, but this would no longer provide a relative approximation error guarantee on~$\log\det{\mat A}$.
Theorem~4.1 by \citet{Ubaru2017fastTrA} is not applicable, since the log of the eigenvalues of the kernel matrix can be of different sign.
Theorem~2 by \citet{Gardner2018gpytorch} is a consequence of Theorem~4.1 by \citet{Ubaru2017fastTrA} and therefore also not applicable. %
\citet{Gardner2018gpytorch} recommend certain default parameter values, though we observed experimentally that this configuration yields estimates whose relative errors are \emph{more often than not} worse than $0.1$ and may vary over two orders of magnitude (see \cref{fig:ste} in \cref{app:results}).
We therefore did not compare our approach to their method.
To nevertheless allow the reader to assess the difficulty of the numerical problems considered in \cref{sec:experiments}, we compare our method to the pivoted Cholesky decomposition of \citet{Harbrecht2012pivotedCholesky} (see \cref{sec:baseline}).

Most related to our \cref{thm:determinant_stopping} is the work by \citet{Mnih2008EBstopping} and references therein. 
They propose an algorithm called \pname{EBStop} that returns an estimate of the mean of a sum of i.i.d.~random variables.
\cref{thm:determinant_stopping} is more general and assumes only a (non-strict) decrease in conditional expectation.
Their approach is in a sense more sophisticated as they also monitor the empirical variance of the addends, which is future work for our us.

\subsection{Cholesky decomposition}
In the following, we will focus on an implementation of the Cholesky decomposition that proceeds \emph{row-wise} over the elements of the matrix, \cref{algo:chol_stopped}.
As opposed to a column-wise or submatrix implementation, the number of floating operations increases with each iteration of the outer loop \parencite{George1986parallelCholesky}.
Hence, this version can benefit the most from early stopping.
\cref{algo:chol_stopped} is useful to express and motivate our idea.
To exploit blocking and parallel computation resources requires some modifications which we describe in \cref{app:practical_implementation}.
Note that computing $\mat C_{jj}$ requires access only to the first $\vec x_1 ,\ldots, \vec x_j$ datapoints. 

\begin{algorithm}
\caption{Augmented row-wise Cholesky decomposition with optional stopping. 
Highlighted are our modifications to the original algorithm.}
\label{algo:chol_stopped}
\begin{algorithmic}[1]
	\State Given $\mat A$, $N$, \augmentedStatement{$\sigma^2$ and $\lBound^+\geq \log\left(\max_{j}\mat A_{jj}\right)$}
	\AState{$D\gets 0, c_\delta\gets (\lBound^+-\log(\sigma^2))H_N^{-1}(\nicefrac{\delta}{2})$}
	\For{$\indk{}=1,\dots,N$}
	\For{$\indi{}= 1, \dots, \indk{}-1$}
	\For{$\indj{}=1,\dots,  \indk{}-1$}
	\State {$\mat A_{\indi{}\indk{}}\gets \mat A_{\indi{} \indk{}}-\mat A_{i\indj{}}\mat A_{\indk{}\indj{}}$}
	\EndFor
	\State {$\mat A_{\indi{}\indk{}}\gets \nicefrac{\mat A_{\indi{}\indk{}}}{\mat A_{\indk{}\indk{}}}$} \hfill\COMMENT{now $\mat A_{\indi{}\indk{}}=\mat C_{\indi{}\indk{}}$}
	\EndFor
	\For{$\indj{}=1, \dots,   \indk{}-1$}
	\State {$\mat A_{\indk{}  \indk{}}\gets \mat A_{  \indk{}  \indk{}}-\mat A_{\indk{}\indj{}}\mat A_{  \indk{}\indj{}}$}
	\EndFor
	\State {$\mat A_{\indk{}  \indk{}}\gets \sqrt{\mat A_{  \indk{}  \indk{}}}$} \hfill\COMMENT{now $\mat A_{\indk{}  \indk{}}=\mat C_{\indk{}\indk{}}$}
	\AState{$D\gets D + 2\cdot \log(\mat A_{\indk{}\indk{}})$}
	\AState {$\hat{D}\gets$ \textbf{EvaluateConditionsAndEstimator}($N, n, D,\sigma^2, c_\delta, \lBound^+$)}
	\IF {\colorbox{newcode}{$\hat{D}\neq 0$}}
		\State{\textbf{return} \colorbox{newcode}{$\hat{D}$}}
	\ENDIF
	\EndFor
	\State{\textbf{return} \colorbox{newcode}{$D$}} \hfill\COMMENT{Now the lower-triangular part of $\mat A$ contains $\mat C$.}
\end{algorithmic}
\end{algorithm}

\section{Stopped Cholesky Decomposition}
\label{sec:method}

This section is a high-level description of our algorithm.
The formal proof of our claims is deferred to Section \ref{sec:main} and the supplementary material.
The main idea of the algorithm is the following: 
each time a new diagonal element of the Cholesky decomposition is computed, we compute an upper bound and a lower bound of $\log\det{\mat A}$.
If the two bounds are sufficiently close to each other and sufficiently far away from zero, a certain relative error can be guaranteed.
We first introduce the bounds used by our algorithm, and then define more precisely what we mean by ``close.''

Denote by~$n$ the number of diagonal elements that have been computed so far.
Our lower bound $\L_n$ is deterministic. 
It is simply the sum of log of the elements computed so far: $\lDetn{n}\ce 2\sum_{j=1}^n \log \mat C_{jj}$, plus a linear extrapolation in~$\sigma^2$. 
That is,
\[
\L_n \ce \lDetn{n}+(N-n)\log \sigma^2
\, .
\]
On the other hand, the upper bound is probabilistic. 
We show in Section~\ref{sec:main} how we can achieve the control of the failure probability.
The key observation is that the diagonal elements of the Cholesky \textbf{decrease in (conditional) expectation}, under the assumption that $\vec x_1,\ldots, \vec x_N$ are independent and identically distributed. (This assumption is not always fulfilled, \eg{}, when the inputs are sorted.
However, in practice, the assumption can be considered established, after a random shuffle of the dataset.) 
The intuition is that, for kernel matrices, one can write
\begin{equation}
\label{eq:key_observation}
\mat C_{nn}^2=k(\vec x_n, \vec x_n)+\sigma^2-\vec k_n\Trans\inv{(\mat K_{n-1}+\sigma^2\mat I_{n-1})}\vec k_n
\, ,
\end{equation}
where $\vec k_n\Trans\ce [k(\vec x_n, \vec x_1), \ldots, k(\vec x_n, \vec x_{n-1})]\Trans\in \Re^{n-1}$.
Hence the diagonal elements of the Cholesky, squared, correspond to the posterior variance of a Gaussian process given observations disturbed by independent Gaussian noise (see \citet[p.~16]{RasmussenWilliams}).
With increasing~$n$, this variance can only decrease.
Thus, the mean of all $\mat C_{nn}$ is likely to be an overestimate of the expected value of $\mat C_{n+1, n+1}$.
Therefore, we use as an upper bound, the sum of the elements computed so far, plus a linear extrapolation of their mean: 
\[
\U_n' \ce \lDetn{n} + (N-n)\frac{\lDetn{n}+c_\delta}{n}+c_\delta
\, ,
\]
where $c_\delta$ depends on the desired failure probability $\delta$.
We defer the exact expression of $c_\delta$ to \cref{sec:main}.
A \emph{deterministic} upper bound to $\log\det{\mat A}$ is
\begin{align}
\label{eq:bound_deterministic}
\U''_n \ce \lDetn{n} + (N-n)\log\left(\sigma^2+\max_{j\in \{1, ..., N\}}k(\vec x_j, \vec x_j)\right)
\end{align}
which is a consequence of Lemma~\ref{lemma:boundedness}.
To make sure that our bound is never worse than this deterministic bound we set 
\begin{align}
\label{eq:bound_final_form}
\U_n\ce \min(\U'_n, \U''_n)\, .
\end{align}
Now we are nearly ready to write our algorithm. 
The only missing piece is to decide whether $\U_n$ and $\L_n$ are close enough. 
Suppose we believe that $\log\det{\mat A}\in [\L_n, \U_n]$, then $\hat{D}_n\ce\frac{1}{2}(\U_n+\L_n)$ is a natural estimate.
We will show (Lemma~\ref{lemma:rel_err_bound}) that it is possible to guarantee the \emph{relative precision}
\begin{align}
\label{eq:relative_precision}
\abs{\frac{\log\det{\mat A} - \hat{D}_n}{\log\det{\mat A}}}\leq r
\, ,
\end{align}
when $\log\det{\mat A}$ cannot be zero, and 
\begin{align}
\label{eq:condition_relative_one}
\stopTimeTwo{}.
\end{align}
To exclude $\log\det{\mat A}=0$, we check in addition that 
\begin{align}
\label{eq:condition_sign_one}
\sign(\L_n)=\sign(\U_n)\neq 0\, .
\end{align}
\cref{algo:main} describes above elaborations in pseudo code.
\cref{algo:chol_stopped} shows our modifications with new statements highlighted.
Importantly, the computation of the bounds and checks are inexpensive in comparison to an outer-loop iteration of the Cholesky decomposition.
\cref{fig:results_bound_early_stop,fig:results_bound_late_stop} show the progression of \cref{eq:condition_relative_one} for two examples.
Note that when $\X$ is bounded and the kernel is differentiable, with a sufficient amount of data, the upper bound gets arbitrarily close to the lower bound.

\begin{algorithm}
	\caption{\label{algo:main}EvaluateConditionsAndEstimator. At a given step, this routine computes the lower and upper bounds, and proceeds to check if they are close enough. 
	}
	\begin{algorithmic}[1]
		\STATE Given $N, n, D_n, \sigma^2$, $c_\delta$ and $\lBound^+$
		\STATE $\L_n\gets D_n+(N-n)\log\sigma^2$
		\STATE $\U_n\gets \min(D_n+(N-n)\frac{D_n+c_\delta}{n}+c_\delta, D_n+(N-n)\lBound^+$
		\IF{$\sign(\U_n)=\sign(\L_n)\neq 0$ and $\U_n-\L_n< 2r\min(\abs{\U_n}, \abs{\L_n})$}
		\STATE \textbf{return} $\frac{1}{2}(\U_n+\L_n)$
		\ENDIF
		\STATE \textbf{return} 0
	\end{algorithmic}
\end{algorithm}

\begin{figure}
	{\centering\ \ \ref{leg:bound}\\}%
	\setlength{\figwidth}{.5\textwidth}%
	\setlength{\figheight}{\figwidth}%
	\setlength{\tabcolsep}{-3pt}%
	\begin{tabular}[t]{lr}
		\renewcommand{\labelx}{processed datapoints}%
		\renewcommand{\labely}{relative error bound}%
		\renewcommand{\ticklabelsy}{true}%
		\hidelegend{}%
		\renewcommand{\runinfo}[7]{\textbf{#1} (D=#2)}%
		\adjustbox{valign=t}{\input{tikz/OU/bank/bound_1.0.tikz}}%
		&
		\renewcommand{\runinfo}[7]{}%
		\renewcommand{\labelx}{$\frac{t}{\operatorname{mean}(t_{\text{default}})}$}%
		\renewcommand{\labely}{~}%
		\adjustbox{valign=t}{\input{tikz/OU/bank/inverse_plot_1.0.tikz}}%
	\end{tabular}
	\caption{a disadvantageous scenario.
		We compute the log-determinant of a kernel matrix using the OU kernel ($\theta=1$, $\ell=\exp(1)$ in Eq.~\eqref{eq:kernel_ou}) on the \BANK{} dataset for ten random permutations.
		\\
		\textbf{Left panel:} same setup as in \cref{fig:results_bound_early_stop}.
		On this dataset, even using a long length-scale, requires processing more than 90\% of the data to achieve a relative error $r$ of at least 0.1.
		\\
		\textbf{Right panel:} same setup as in \cref{fig:results_bound_early_stop}.
		When our algorithm is not stopping early, that is, it returns the result of the default Cholesky, the overhead is on average less than 5\%.
		The Cholesky with pivoting on the other hand may require more than 150\% of the time of the default Cholesky.
		The extreme difference in absolute runtime between this figure and \cref{fig:results_bound_early_stop} is investigated in \cref{sec:result_description}.
	}
	\label{fig:results_bound_late_stop}
\end{figure}

\section{Theoretical Justification}
\label{sec:main}

We now turn to the theoretical analysis of our algorithm. 
Our main goal in this section is to explain how the expressions of the lower and upper bounds are obtained. 
Note that we consider, in fact, a more general problem: stopping the computation of a sum of random variables that decrease in expectation.
To the best of our knowledge, this is the first result obtained in this setting, where the addends are not independent and identically distributed (the $\vec x_i$ are not the addends). 
Theorem~\ref{thm:determinant_stopping} states that the stopping condition described in the following is a solution to this problem, and Theorem~\ref{thm:determinant_stopping_application} states that Theorem~\ref{thm:determinant_stopping} can be applied to estimate determinants of kernel matrices.

\subsection{Notation}
Since we are considering an optional stopping problem, we need to use the terminology of stochastic processes. 
This section is a quick reminder of the most important concepts, we refer to \citet{Grimmett2001RandomProcesses}, and \citet{Davidson1994StochasticLimitTheory} for a more thorough introduction. 
For a monotonically increasing function $f:\Re\rightarrow \Re$ and $\delta\in \Re$, define $f^{-1}(\delta)\ce \arg\sup_{\epsilon \in \Re}\{f(\epsilon)\leq \delta\}$.
A \emph{filtration} is a sequence $(\mathcal{F}_j)_{j\in\mathbb{N}}$ of increasing $\sigma$-algebras, \emph{i.e.}, $\mathcal{F}_{j}\subseteq\mathcal{F}_{j+1}$ for all $j\in\mathbb{N}$.
For random variables $X_1, \ldots, X_N$, we denote by $\sigma(X_1, \ldots, X_N)$ the $\sigma$-algebra generated by $(X_1,\ldots,X_N)$.
A sequence of random variables $(X_j)_{j\in \mathbb{N}}$ is called \emph{adapted} to a filtration, if $X_j$ is $\mathcal{F}_j$-measurable for all $j\in \mathbb{N}$.
A random variable $\tau$ is called a \emph{stopping time} (w.r.t.~a filtration), if it takes values in $\mathbb{N}$ and $\{\tau=j\}\in\mathcal{F}_j$ for all $j\in\mathbb{N}$.

\begin{table*}
	\begin{tabular}{l|ccp{0.65\textwidth}}
		Key & $N$ & $D$ & Source \& URL 
		\\\hline
		\\\BANK{} & 45211 & 51 & \textcite{Moro2014BankMarketingDataset}
		\\&&& \url{Bank+Marketing} 
		\\\METRO & 48204 & 66 & no citation request\\
		&&& \url{Metro+Interstate+Traffic+Volume}
		\\\PM & 43824 & 79 & \citet{Liang2015pmDataset}\\
		&&& \url{Beijing+PM2.5+Data}
		\\\PROTEIN & 45730 & 9 & no citation request\\
		&&&\url{Physicochemical+Properties+of+Protein+Tertiary+Structure}
		\\\PUMADYN & 8192 & 32 & \citet{Snelson2006pseudo}\\ &&&\url{www.cs.toronto.edu/~delve/data/pumadyn/desc.html}
		\\\TAMILNADU\footnote{With our implementation we were unable to load this dataset in its original state. We removed all \' and superfluous blanks.} & 45781 & 53 & no citation request\\
		&&& \url{Tamilnadu+Electricity+Board+Hourly+Readings}
	\end{tabular}
	\caption{
		Overview over all datasets used for the experiments in Section~\ref{sec:experiments}. 
		Key refers to the title, we gave a dataset in this article.
		The letter~$N$ refers to the number of instances (training and testing) and~$D$ refers to the dimensionality after one-hot encoding. 
		The URL is a suffix for \url{http://archive.ics.uci.edu/ml/datasets/}.
		The reference in Source acknowledges a citation request, if any.  %
	}
	\label{tbl:datasets}
\end{table*}

\subsection{Problem Setting}
\label{sec:empirical_estimator_problem_definition}
Let $(\Omega, \mathcal{F}, \Proba)$ be a probability space and $(\mathcal{F}_j)_{j\in \{1,\ldots,N\}}$ be a filtration. 
Furthermore, let $(f_j)_{j\in \{1,\ldots,N\}}\in [\lBound^-, \lBound^+]$ be a sequence of random variables such that for $j \in \{1, \ldots, N-1\}:$ $f_j$ is $\mathcal{F}_j$-measurable and the \textbf{conditional expectation is decreasing}, formally: 
\begin{equation}
\label{eq:det_crucial_assumption}
\Exp[f_{j+1} \mid \mathcal{F}_j]\leq \Exp[f_{j}\mid \mathcal{F}_{j-1}] \tag{$\boldsymbol\ast$}
\, ,
\end{equation}
with $\mathcal{F}_0\ce\{\emptyset, \Re\}$. For this sequence, we want to estimate its sum
\begin{equation}
\label{eq:determinant_sum}
D_N\ce\sum_{j=1}^N f_j
\, .
\end{equation}
Given a desired upper bound on the relative error $r\in (0, 1)$ and a probability of failure $\delta\in(0, 1)$, our goal is to device a strategy that, being presented sequentially with the $f_1, f_2,\ldots$, decides in each step whether to continue or to stop, and if stopping, provides an estimator $\hat{D}_\tau$, such that its relative error is less than $r$ with probability $1-\delta$.
Formally, the goal is to device a stopping time $\tau$ and an estimator $\hat{D}_\tau$, such that, 
\begin{equation}
\label{eq:determinant_requirement}
\proba{\abs{\frac{D_N-\hat{D}_\tau}{D_N}}>r} \leq \delta
\, .
\end{equation}
\begin{remark}
	A trivial solution is to define $\tau\ce N$ and $\hat{D}_\tau\ce D_N$, which simply consists in doing the whole computation.
\end{remark}

\subsection{Stopping Condition}
\label{sec:det_stopping_condition}

We now define precisely the quantities introduced in \cref{sec:method}: the lower bounds~$\L_n$ and the upper bounds~$\U_n$. 
Recall that the lower bounds~$\L_n$ are deterministic, whereas~$D_N\leq \U_n$ holds only with a certain probability.
The stopping time~$\tau$ will monitor these bounds and stop if they are large in magnitude (away from zero) and close enough that the relative error cannot exceed the desired precision~$r\in (0,1)$.

As in \cref{sec:method}, set 
\begin{align}
\L_n&\ce D_n+(N-n)\lBound^-, \label{eq:lower_bound}
\\\U_n&\ce D_n +\min\left(c_\delta+(N-n)\frac{D_n+c_\delta}{n}, (N-n)\lBound^+\right), \label{eq:upper_bound}
\\\hat{D}_n&\ce \frac{1}{2}(\L_n+\U_n), \label{eq:det_estimator}
\end{align}
where $c_\delta\ce (\lBound^+-\lBound^-) \inv{H_N}(\nicefrac{\delta}{2})$ and 
\[
H_N(x)\ce \one_{\{x\leq N\}}\sqrt{\left(\frac{N}{N+x}\right)^{N+x}\left(\frac{N}{N-x}\right)^{N-x}}.
\]
The function $H_N$ is derived from a theorem by \citet{Fan2012Hoeffding} which our proofs rely on.

Finally, we define the stopping time as
\begin{equation}
\label{eq:stopping_time}
\tau = N \wedge \min\bigl\{n < N \text{ s.t. } C_n^s \text{ and } C_n^p\text{ hold}\bigr\}
\, ,
\end{equation}
where $C_n^s$ is the \emph{sign condition}
\begin{equation}
\label{eq:condition_sign}
C_n^s \enspace\text{true if}\enspace \stopTimeOne
\, ,
\end{equation}
and $C_n^p$ is the \emph{relative precision} condition
\begin{equation}
\label{eq:condition_relative}  %
C_n^p \enspace\text{true if}\enspace \stopTimeTwo
\, .
\end{equation}
Note that the quantities in the stopping conditions are all $\mathcal{F}_n$-measurable, thus $\tau$ is indeed a stopping time. 
We can now state our main result. 

\begin{theorem}
	\label{thm:determinant_stopping}
	Assume that $D_N$ is a sum of random variables decreasing conditionally in expectation as in \cref{sec:empirical_estimator_problem_definition}. 	
Then, for any $r,\delta\in (0,1)$, the relative error of the estimator $\hat{D}_\tau$ defined by \cref{eq:lower_bound,eq:upper_bound,eq:stopping_time,eq:condition_sign,eq:condition_relative} is bounded by $r$ with probability at least $1-\delta$, formally:
\begin{align}
\proba{\abs{\frac{D_N-\hat{D}_\tau}{D_N}}>r} \leq \delta\, .\notag 
\end{align}
\end{theorem}
Intuitively, \cref{thm:determinant_stopping} guarantees that stopping early in the computation makes sense for any given~$r$ and~$\delta$. 
The less precision is required (corresponding to larger $r$) the easier the second stopping condition in \cref{eq:condition_relative} can be satisfied.
The less confidence is necessary (corresponding to larger $\delta$), the smaller the term $c_{\delta}$ in \cref{eq:upper_bound}, which also increases chances to satisfy \cref{eq:condition_relative} earlier.
On the other hand, when $r=0$, \cref{eq:upper_bound} can only be true, if upper and lower bounds coincide.
The latter can only be the case if $c_{\delta}=0$ (requires $\delta=2$) and $D_n=n\lBound^-$.
This means: if we were to desire absolute precision, the theorem would recommend to compute the full sum.

The proof of \cref{thm:determinant_stopping}, and the proof the following lemma are part of the supplementary material.
Let us give a sketch of the proof. 
The design of the stopping condition is based on the following Lemma~\ref{lemma:rel_err_bound_main_text}.

\begin{lemma}
	\label{lemma:rel_err_bound_main_text}
	Let $D \in [\L, \U]$, and assume $\sign(\L)=\sign(\U)\neq 0$. 
	Then 
	\[
	\frac{\smallabs{D-(\U+\L)/2}}{\abs{D}}\leq \frac{\U-\L}{2\min(\abs{\L}, \abs{\U})}
	\, .
	\]
\end{lemma}

The proof of \cref{thm:determinant_stopping} first bounds $\proba{\abs{\frac{D_N-\hat{D}_\tau}{D_N}}>r}$ by $\proba{\abs{\frac{D_N-\hat{D}_\tau}{D_N}}>r, D_N\leq \U_\tau}+\proba{D_N>\U_\tau}$.
Using Lemma~\ref{lemma:rel_err_bound_main_text} and the stopping conditions, the probability of the term left of the sum is 0.
We bound $P(D_N>\U_\tau)$ by applying \citet{Fan2012Hoeffding}'s \emph{Hoeffding's inequality for martingales} twice.
Once, to show that $D_N$ is probably not much larger than its expected value, and a second time, to show that $\U_\tau$ is probably not much smaller. 

\subsection{Application to Kernel-Matrix Determinant Estimation}
\label{sec:theorem_application}
We now specialize \cref{thm:determinant_stopping} to the situation at hand. 

\begin{theorem}
	\label{thm:determinant_stopping_application}
	Assume $\vec x_1, \dots, \vec x_N \in \X$ are independent and identically distributed. 
Denote with $\Proba$ the law of the $\vec x_1,\ldots, \vec x_N$ and with $\mat C$ the Cholesky decomposition of $\mat A$.
Define the probability space $(\X, \sigma(\vec x_1, \ldots, \vec x_N), \Proba)$ and the canonical filtration $\mathcal{F}_j\ce \sigma(\vec x_1,\dots, \vec x_j)$ for $j=1, \ldots, N$.
Further, define 
\begin{align}
	f_j&\ce 2\log {\mat C_{jj}}, \notag
	\\\lBound^-&\ce \log \sigma^2 \, , \notag
\end{align}
and assume there exists a constant $\lBound^+$ such that 
\[
\max_{j=1,\dots,N}\log (k(\vec x_j,\vec x_j)+\sigma^2)\leq \lBound^+
\quad\text{ almost surely.}	
\, .
\]
Then, using the definitions of Theorem~\ref{thm:determinant_stopping}, 
\[
\proba{\frac{\abs{\log\det{\mat A}-\hat{D}_\tau}}{\abs{\log\det{\mat A}}}>r} \leq \delta
\, .
\]
\end{theorem}

As stated before, the i.i.d.~assumption is not too stringent. 
Finding the deterministic upper bound $\lBound^+$ is also a given in most use-cases, for example when $\mathbb{X}$ is bounded, or when the kernel is normalized or stationary.
For instance, $\lBound^+=\theta$ in the case of the RBF and OU kernels in \cref{eq:kernel_rbf,eq:kernel_ou} respectively.

The proof of Theorem~\ref{thm:determinant_stopping_application} is part of the supplementary material.
Essentially, to apply Theorem~\ref{thm:determinant_stopping} for the estimation of kernel-matrix determinants, one has to show that the summands are decreasing in expectation. 
As stated before, the key observation is that the diagonal elements of the Cholesky correspond to the posterior variance of a Gaussian process given observations disturbed by independent Gaussian noise.
With each observation, the posterior variance can only decrease, which in turn allows to show that the diagonal elements of the Cholesky decrease conditionally in expectation.

\section{Experiments}
\label{sec:experiments}
\label{sec:stopping_experiments}

One application of our implementation is to probe \emph{bad} kernel parameters quickly. %
For example, consider the case of a kernel matrix generated from an radial basis function (RBF) kernel 
\begin{align}
	k_{RBF}(\vec x, \vec z)&\ce \theta \exp\left(-\frac{\norm{\vec x - \vec z}^2}{2\ell^2}\right) \label{eq:kernel_rbf}
\end{align}
with a lengthscale $\ell$ far too large with respect to the data. 
In that case, the diagonal elements of the Cholesky then come quickly close to $\sigma^2$, which implies that upper and lower bounds become close enough to stop the computation earlier. 
We examine this hypothesis for the RBF on different datasets increasing the length scale exponentially.
Furthermore, to also explore the limitations of our approach, we run the same experiments for the Ornstein-Uhlenbeck (OU) kernel
\begin{align}
	k_{OU}(\vec x, \vec z)&\ce \theta \exp\left(-\frac{\norm{\vec x - \vec z}}{\ell}\right). \label{eq:kernel_ou}
\end{align}
Both kernels are (in the limit) members of the Mat\'ern class of covariance functions \citep[p.~85]{RasmussenWilliams}. 
Whereas samples from a Gaussian process with RBF covariance are the smoothest in this class, samples from the OU are the roughest.
It is therefore not a surprise that our approach is less successful when using the OU kernel.
Though, it is an advantage, that one can quite predict, when stopping is possible or not.

\subsection{Experiment setup}
From the UCI machine learning repository \citep{Dua2019uci}, we took all multivariate datasets in matrix format with $40.000$ to $50.000$ instances without missing values.
Furthermore, we included the frequently used \PUMADYN{} dataset \citep{Snelson2006pseudo} as a small-scale example of only 8000 instances.
Categorical variables where one-hot encoded and each dataset was then standardized.
Table~\ref{tbl:datasets} provides an overview of all the datasets that we use.

All large-scale experiments ($\geq 40.000$ datapoints) were executed on machines running Ubuntu 18.04 with 32 Gigabytes of RAM and two Intel Xeon E5-2670 v2 CPUs.
The experiments for the \PUMADYN{} dataset were run on a laptop running Ubuntu 20.04 with 16 Gigabytes of RAM and an Intel i7-8665U CPU, to demonstrate the usefulness of our approach on more standard hardware.
We remark again that we do not use \cref{algo:chol_stopped} but \cref{algo:chol_blocked} in \cref{app:practical_implementation} which is a more practical implementation capable of exploiting blocking and parallelization.

\subsection{Baseline}
\label{sec:baseline}
As baseline, we compare against the Cholesky decomposition with full pivoting \citep{Harbrecht2012pivotedCholesky}.
In each step $n$, the algorithm keeps track of the approximation error of all remaining diagonal elements $i$---that is how much $\mat K_{ii}$ differs from $[\mat L_n\mat L_n\Trans]_{ii}$---and processes the element inducing the most error next.
The algorithm stops when a certain absolute error tolerance on the diagonal elements can be guaranteed.
Note that $S$ iterations of the pivoted Cholesky require $\O(NS^2)$ operations whereas \cref{algo:chol_stopped} scales as $\O(S^3)$.
For this algorithm, we can set $\U_n^P\ce D_n+\sum_{j=n+1}^N \log (\mat K_{jj}-[\mat L_n\mat L_n\Trans]_{jj})$ and $\L_n^P\ce L_n$, and apply the same stopping strategy which allows to compare this algorithm with our proposed approach.
In the next paragraph, we describe how to compare both algorithms without modifying the Fortran implementation of the Cholesky with pivoting.

\subsection{Parameters and performance metric}
We set $\sigma^2\ce 0.001$ and $\theta\ce 1$, and increased the lengthscale as $\ell\ce \exp(i)$ for $i=-1,\ldots, 3$.
The Cholesky decomposition with full pivoting takes as input parameter a desired relative precision on the diagonal elements $d$ (instead of a relative precision on the log-determinant).
We ran this algorithm for $d\in \{0.001, 0.005, 0.01, 0.05, 0.1, 0.5\}$.
After the pivoted Cholesky stopped, we computed the relative precision on the log-determinant that this algorithm could guarantee in that step.
Then we ran our algorithm trying to achieve the same precision for $\delta=0.1$. %
Occasionally, the desired relative precision is larger than $1$.
In that case, $\hat{D}_N\ce 0$ is an estimator satisfying this requirement which would allow stopping before even starting.
However, we did not check for this condition, to instead observe when the algorithm would stop in such situations.
We repeated each configuration for ten random permutations of the dataset.
We measured the performance of our method in terms of CPU time $t$ saved over the average CPU time used for the default Cholesky $t_{\operatorname{default}}$:
\begin{align}
\metric \ce \frac{t}{\operatorname{mean}(t_{\operatorname{default}})}.
\end{align}
Thus, small values of $\metric$ are better. %

\subsection{Results}
\label{sec:result_description}
As an example, \cref{fig:results_time_pm} shows our results for the \PM{} dataset.
For all other datasets, similar figures (\cref{fig:results_time_protein,fig:results_time_tamilnadu_electricity,fig:results_time_bank,fig:results_time_metro,fig:results_time_pumadyn}) can be found in \cref{app:results}.
In all experiments, the returned estimate of our modified Cholesky decomposition had indeed the desired precision.

For the easy cases, our algorithm needs less than $10\%$ of the average time of the default Cholesky.
Here, with easy cases we mean that the relative error can be larger than $0.1$ and using an RBF kernel with $\ell \geq \exp(1)$ (there is one exception: the \BANK{} dataset and $\ell=\exp(1)$).
The Cholesky decomposition with pivoting also saves time in these settings, yet less.
The difference between the algorithms becomes more apparent the harder the problem.
Except for three cases, which we will elaborate below, our algorithm needs never longer than $105\%$ of the time of the default Cholesky.
In contrast, the Cholesky with pivoting may take more than twice as long.

In three cases our approach crosses the $105\%$ mark: using an RBF kernel with $\ell=1$ on \PM{} and \METRO{}, and $\ell=\exp{(1)}$ on \METRO{}.
In these scenarios, the kernel matrix contains many extremely small entries of less than $10^{-65}$.
Floating-point multiplication is not a constant operation and we observed that a large number of such entries significantly prolongs the runtime of our experiments.
It is the reason why for $\ell=\exp(-1)$, the run time for the default Cholesky can take up to ten times longer than for larger length-scales.
Our row-wise implementation of the Cholesky decomposition suffers \emph{more} from this phenomenon than the original OpenBLAS version.
One can circumvent this problem by eliminating such small entries or by increasing the block-size in \cref{algo:chol_blocked}.
However, we deliberately did not apply these strategies to showcase possible downsides of this implementation.
Furthermore, note that the absolute overhead is less than 30s for these three cases and that the effect becomes more negligible the longer the absolute running time.
Importantly, the additional run-time does not stem from checking our stopping conditions.
\renewcommand{\ticklabelsy}{true}%
\renewcommand{\labelx}{$\substack{d\\r}$}%
\renewcommand{\labely}{$\metric$}%
\setlength{\figwidth}{.48\textwidth}%
\setlength{\figheight}{.16\textheight}%
\begin{figure}
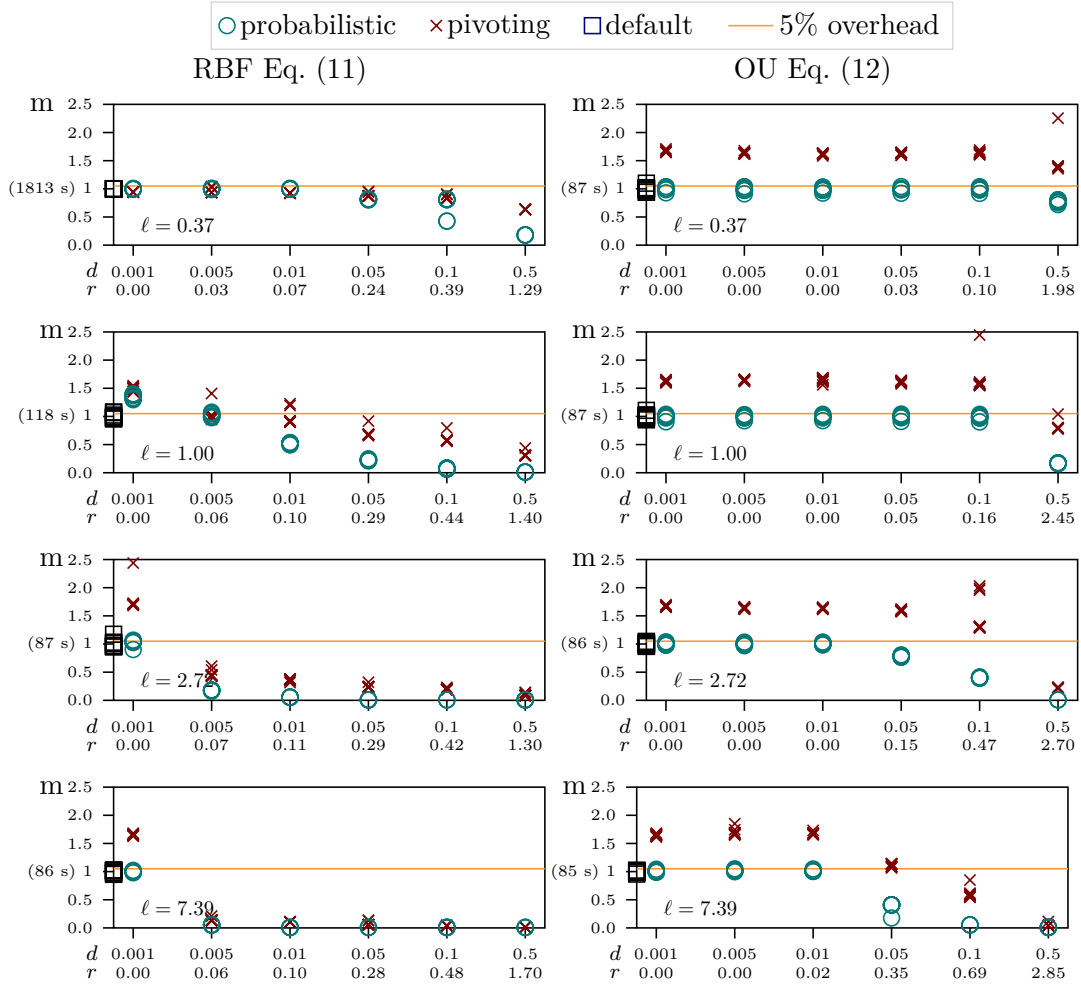
%
{\centering\ref{leg:performance}\\}
\setlength{\tabcolsep}{-4pt}
\begin{tabular}{rr}
\multicolumn{1}{c}{RBF \cref{eq:kernel_rbf}}&\multicolumn{1}{c}{OU \cref{eq:kernel_ou}}\\
\renewcommand{\labely}{\metric}%
\input{tikz/RBF/pm25/joint_plot_-1.0.tikz}
&
\renewcommand{\labely}{\metric\ \ }%
\input{tikz/OU/pm25/joint_plot_-1.0.tikz}\\
\renewcommand{\labely}{\metric\ }%
\input{tikz/RBF/pm25/joint_plot_0.0.tikz}
&
\renewcommand{\labely}{\metric\ \ }%
\input{tikz/OU/pm25/joint_plot_0.0.tikz}\\
\renewcommand{\labely}{\metric\ \ }%
\input{tikz/RBF/pm25/joint_plot_1.0.tikz}
&
\renewcommand{\labely}{\metric\ \ }%
\input{tikz/OU/pm25/joint_plot_1.0.tikz}\\
\renewcommand{\labely}{\metric\ \ }%
\input{tikz/RBF/pm25/joint_plot_2.0.tikz}
&
\renewcommand{\labely}{\metric\ \ }%
\input{tikz/OU/pm25/joint_plot_2.0.tikz}
\end{tabular}
\caption{
relative execution times to compute the log-determinant using RBF (\textbf{left panel}) and OU (\textbf{right panel}) kernels on the \PM{} dataset for $\theta=1$, $\log\ell=-1, \dots, 3$ and $\delta=0.1$ for ten repetitions.
The number next to one on the $y$-axis displays the absolute execution times of the default Cholesky.
The solid, horizontal, orange line (\ref{leg:overhead}) visualizes the $105\%$ mark.
The $x$-axis displays a desired absolute precision on the diagonal elements $d$ (top) and the average corresponding desired relative precision~$r$ (bottom) on the log-determinant.
The longer the length-scale, the earlier it is possible to stop and the higher the speed-up.
The speed-up is generally higher for the RBF than for the OU.
Even though the Cholesky decomposition with pivoting (\ref{leg:PivotedChol}) needs to compute less diagonal elements (see \cref{fig:results_bound_early_stop,fig:results_bound_late_stop}) compared with our methods it is slower in practice and may even take more than twice as long as simply running the default Cholesky decomposition (\ref{leg:DefaultChol}).
Our method (\ref{leg:StoppedChol}) on the other hand is faster, and when approximation is hard, the overhead is negligible. 
One exception can be seen for the RBF kernel and using a length-scale of $\ell=1$.
The reason for this exception is described in \cref{sec:result_description}.
}
\label{fig:results_time_pm}
\end{figure}

\section{Conclusion}
\label{sec:future}

\subsection{Summary}
We presented a stopping strategy for the Cholesky decomposition that allows to obtain estimates for the log-determinant of a kernel matrix of desired precision $r$, before completing the decomposition.
The stopping strategy has only one parameter: a failure probability $\delta$.
We showed that the returned estimate has this desired precision with probability $1-\delta$, under the mild assumptions that the dataset inputs are independent and identically distributed and a boundedness assumption that is met if the kernel or the domain is bounded. 
We demonstrated that there exists settings in which it is possible to save considerable amounts of time when stopping the Cholesky decomposition before completion.
Importantly, when not stopping early, the induced overhead is less than five percent on average.

As part of their concluding remarks, \citet{Chalupka2013comparison} wrote that
\begin{quotation}
	\begin{quote}
		...the results presented above point to the very simple Subset of Data method (or the Hybrid variant) as being the leading contender. We hope this will act as a rallying cry to those working on GP approximations to beat this ``dumb'' method.
	\end{quote}
\end{quotation}
In essence, the presented idea makes a virtue of necessity.
Algorithm \ref{algo:main} can be viewed as an estimate for how much data is necessary to identify a kernel model for a particular dataset distribution.
The claim that kernel machines do not scale well with large datasets becomes brittle, when the overall dataset size matters little.

\subsection{Future work}
Early stopping for lower precision values $r$, closer to numerical precision would be desirable.
One way to achieve this goal could be to find a less conservative, \emph{probabilistic} lower bound on the log determinant.
A direction to investigate are concentration inequalities for self-bounding functions \parencite[p.~60]{Boucheron2013concentration}.
Some concentration inequalities for self-bounding functions allow to reason about the probability of the function falling \emph{below} its expectation.
One can show, that the log-determinant of a kernel matrix is such a function.

In the long run, we hope to lift our experiments to hyper-parameter optimization for Gaussian processes.
For that, our analysis needs to be extended to the term $\vec y\Trans\inv{\mat A}\vec y$.
This analysis is similar, but not trivial.
\acks{Funding for this research was provided by the Danish Ministry of Education and Science, Digital Pilot Hub and Skylab Digital.

Simon is grateful for patient listening and fruitful discussions to Gabriele Abbati, Philipp Hennig, Motonobu Kanagawa, Hans Kersting, Jonas K\"ubler, Simon Julien-Lacoste, Krikamol Muandet, Alexander Neitz, Giambatista Parascandolo, Micha\"el Perrot, Carl Rasmussen, Luca Rendsburg, Maja Rudolph, Michael Schober, Sebastian Weichwald and Inna Zeitler.}

\appendix
\crefalias{Section}{Appendix}
\section{A practical implementation of Cholesky decomposition with stopping}
\label{app:practical_implementation}
\begin{algorithm}
	\caption{Blocked and recursive formulation of \cref{algo:chol_stopped}.}
	\label{algo:chol_blocked}
	\begin{algorithmic}[1]
		\State Given $\mat A$, $N$, $b$, \augmentedStatement{$\sigma^2$ and $\lBound^+\geq \log\left(\max_{j}\mat A_{jj}\right)$}
		\AState{$D\gets 0, c_\delta\gets (\lBound^+-\log(\sigma^2))H_N^{-1}(\nicefrac{\delta}{2})$}
		\State $\mat A_{1:b,1:b}\gets \operatorname{chol}(\mat A_{1:b, 1:b})$
		\AState{$D\gets D + 2\cdot \sum_{l=1}^{b}\log(\mat A_{ll})$}
		\AState {$\hat{D}\gets$ \textbf{EvaluateConditionsAndEstimator}($N, b, D,\sigma^2, c_\delta, \lBound^+$)}
		\IF {\colorbox{newcode}{$\hat{D}\neq 0$}}
			\State{\textbf{return} \colorbox{newcode}{$\hat{D}$}}
		\ENDIF	
		\State $i\gets b+1$, $j\gets \min(i+b, N)$
		\While{$i<N$}
			\State $\mat A_{i:j,1:i} \gets \mat A_{i:j,1:i}\mat A_{1:i,1:i}\iTrans$
			\State $\mat A_{i:j,i:j}\gets \mat A_{i:j,i:j} - \mat A_{i:j,1:i}\mat A_{i:j,1:i}\Trans$
			\State $\mat A_{i:j,i:j}\gets \operatorname{chol}(\mat A_{i:j,i:j})$
			\AState{$D\gets D + 2\cdot \sum_{l=i}^{j}\log(\mat A_{ll})$}
			\AState {$\hat{D}\gets$ \textbf{EvaluateConditionsAndEstimator}($N, j, D,\sigma^2, c_\delta, \lBound^+$)}
			\IF {\colorbox{newcode}{$\hat{D}\neq 0$}}
				\State{\textbf{return} \colorbox{newcode}{$\hat{D}$}}
			\ENDIF		
			\State $i\gets i+b$, $j\gets \min(i+b, N)$
		\EndWhile
		\State{\textbf{return} \colorbox{newcode}{$D$}}
	\end{algorithmic}
\end{algorithm}

\cref{algo:chol_blocked} is a blocked and recursive version of \cref{algo:chol_stopped}.
Our OpenBLAS implementation uses the above algorithm with a block size of $b\ce\#CPUS \cdot BLOCK\_SIZE$, where $BLOCK\_SIZE$ is the internal OpenBLAS block size.
Furthermore, the call to \texttt{chol} is a call to the default OpenBLAS Cholesky.
\cref{algo:chol_blocked} is easy to employ in or on top of any library.

\section{Proof of Theorem~\ref{thm:determinant_stopping_application}}
\label{app:proof_thm_application}

\begin{proof}
	By Lemma~\ref{lemma:determinant_from_chol}: $\log\det{\mat A}=\sum_{j=1}^N \mat C_{jj}$, and one can see that the problem already has the right form for (main paper) Theorem~\ref{thm:determinant_stopping}.
	To apply the theorem, we need to show that for all $j=1, \ldots N$, the $\mat C_{jj}$ are functions of $\vec x_1, \ldots, \vec x_j$ (Lemma~\ref{lemma:cholesky_and_gp_variance}), that $f_j\ce 2\log\mat C_{jj}\in[\lBound^-,\lBound^+]$ (Lemma~\ref{lemma:boundedness}), and that $\Exp[f_{j+1}\mid \mathcal{F}_j]\leq \Exp[f_j\mid \mathcal{F}_{j-1}]$ (Lemma~\ref{lemma:decreasing_expectation}).
\end{proof}
We now proceed just as in the proof above.
We are going to show that the $j$-th diagonal element of the Cholesky is bounded and can be computed from $\vec x_1,.., \vec x_j$ only.
Then, we conclude that the elements must decrease in (conditional) expectation.
To proof the following lemmata, define
\begin{align}%
	\kj[n] & \ce [k(\vec x, \vec x_1), \ldots, k(\vec x, \vec x_n)]\Trans\in\Re^n \text{ ,}\\
	\vec k_{n+1} & \ce \vec k_{n}(\vec x_{n+1})\in \Re^n \text{ and}\\
	v_n & \ce k(\vec x_n, \vec x_n) + \sigma^2 - \vec k_n\Trans \inv{(\mat K_{n-1}+\sigma^2\mat I_{n-1})}\vec k_n\, .
\end{align}%
The first term~$\kj[n]$ denotes the covariance between an arbitrary input~$\vec x$ and the first $n$~datapoints from the dataset.
In particular, this definition will be used in the proof of Lemma~\ref{lemma:decreasing_expectation}, which states the decrease in expectation.
The term $v_n$ is the posterior variance of a Gaussian process $f$ conditioned on observations $\vec y\in \Re^n$, perturbed by Gaussian white noise\footnote{see for example \citet[p.~16]{RasmussenWilliams}}: $p(\vec y\mid \vec f)=\N(\vec 0, \sigma^2\vec I)$.
Lemma~\ref{lemma:cholesky_and_gp_variance} establishes a link between $v_n$ and the $n$-th diagonal element of the Cholesky, which is then used in the proof of Lemma~\ref{lemma:decreasing_expectation}.

\begin{lemma}[Link between the Cholesky and Gaussian process regression]
	\label{lemma:cholesky_and_gp_variance}
	Denote with $\mat C_N$ the Cholesky decomposition of $\mat A\ce \mat K_N+\sigma^2\mat I_N$, so that $\mat C_N\mat C_N\Trans=\mat A$.
	The $n$-th diagonal element of $\mat C_N$, squared, is equivalent to $v_n$:
	\[
	[\mat C_N]_{nn}^2=v_n
	\, .
	\]
\end{lemma}

\begin{proof}
	By a slight abuse of notation, let us define 
	\begin{align}%
		\mat C_1&\ce \sqrt{k(\vec x_1, \vec x_1)+\sigma^2}\, ,\quad \text{and} \\
		\mat C_N&\ce \begin{bmatrix}
			\vec C_{N-1} & \vec 0 \\
			\vec k_N\Trans\mat C_{N-1}\iTrans & \sqrt{v_N}
		\end{bmatrix}
		\, .
	\end{align}%
	We will show that the lower triangular matrix $\mat C_N$ satisfies $\mat C_N\mat C_N\Trans = \mat K_N+\sigma^2\mat I_N$.
	Since the Cholesky decomposition is unique \citep[Theorem~4.2.7]{golub2013matrix4}, $\mat C_N$ must be the Cholesky decomposition of $\mat K_N+\sigma^2\mat I_N$.
	Furthermore, by definition of $\mat C_N$, $[\mat C_N]_{NN}^2=v_N$.
	The statement then follows by the recursive definition of $\mat C_N$.	
	
	We want to show that $\mat C_N\mat C_N\Trans = \mat K_N+\sigma^2\mat I_N$. 
	The proof follows by induction.
	To show the beginning, note that
	\begin{align}%
		\mat C_1 \mat C_1\Trans = k(\vec x_1, \vec x_1)+\sigma^2=\mat K_1+\sigma^2\mat I_1\, .
	\end{align}%
	For the induction step, let us assume that the proposition holds up to $N-1$, that is, $\mat C_{N-1}\mat C_{N-1}\Trans=\mat K_{N-1}+\sigma^2\mat I_{N-1}$, then, by definition of $\mat C_N$,
	\begin{align}%
		&\mat C_{N}\mat C_{N}\Trans 
		= \begin{bmatrix}
			\vec C_{N-1} & \vec 0 \\
			\vec k_N\Trans\mat C_{N-1}\iTrans & \sqrt{v_N}
		\end{bmatrix}\cdot\begin{bmatrix}
			\vec C_{N-1}\Trans & \mat C_{N-1}^{\!-1}\vec k_N \\
			\vec 0\Trans & \sqrt{v_N}
		\end{bmatrix}
		\\&=\begin{bmatrix}
			\mat C_{N-1}\mat C_{N-1}\Trans & \mat C_{N-1}\mat C_{N-1}^{\!-1}\vec k_N\\
			\vec k_N\Trans \mat C_{N-1}\iTrans \mat C_{N-1}\Trans &  \vec k_N\Trans\mat C_{N-1}\iTrans\mat C_{N-1}^{\!-1}\vec k_N+v_N
		\end{bmatrix}
		\\&=\begin{bmatrix}
			\mat K_{N-1} +\sigma^2\vec I_{N-1} & \vec k_N\\
			\vec k_N\Trans & \vec k_N\Trans\inv{(\mat K_{N-1}+\sigma^2\mat I)}\vec k_N+v_N
		\end{bmatrix}
		\\&=\begin{bmatrix}
			\mat K_{N-1} +\sigma^2\vec I_{N-1} & \vec k_N\\
			\vec k_N\Trans & k(\vec x_N, \vec x_N)+\sigma^2
		\end{bmatrix} \, .
	\end{align}%
\end{proof}

\begin{lemma}[Bounding the $f_j$s]
	\label{lemma:boundedness}
	Denote by $\mat C_N$ the Cholesky decomposition of $\mat K_N+\sigma^2\mat I_N$. 
	Define $\lBound^{-}\ce \log \sigma^2$ and take $\lBound^+\geq\max_{j=1, \ldots, N} \log\left(k(\vec x_j, \vec x_j)+\sigma^2\right)$. 
	Then, for all $j\in \{1, ..., N\}$, 
	\[
	\lBound^- \leq f_j \leq \lBound^+ \quad \text{a.s.}
	\, .
	\]
\end{lemma}

\begin{proof}
	By Lemma~\ref{lemma:cholesky_and_gp_variance}, $$\mat C_{nn}^2=k(\vec x_n, \vec x_n) + \sigma^2 - \vec k_n\Trans \inv{(\mat K_{n-1}+\sigma^2\mat I_{n-1})}\vec k_n\, .$$ 
	The term $\vec k_n\Trans\inv{(\mat K_{n-1}+\sigma^2\mat I_{n-1})}\vec k_n$ is always positive since $\inv{(\mat K_{n-1}+\sigma^2\mat I_{n-1})}$ is a symmetric positive definite matrix.
	Hence, $k(\vec x_n, \vec x_n)+\sigma^2$ is an upper bound to $\mat C_{nn}^2$.
	On the other hand, since $k$ is a kernel, $k(\vec x_n, \vec x_n)-\vec k_n\Trans\inv{(\mat K_{n-1}+\sigma^2\mat I_{n-1})}\vec k_n$ cannot be negative and~$\sigma^2$ is a therefore a lower bound to $\mat C_{nn}^2$.
	Since both values are positive and the logarithm is an increasing function on the positive real axis, the proof is complete.
\end{proof}

Equipped with the link between the diagonal elements of the Cholesky and Gaussian process regression stated in Lemma~\ref{lemma:cholesky_and_gp_variance}, we can now show that the diagonal elements of the Cholesky must decrease in (conditional) expectation, when treating the $\vec x_1, ..., \vec x_N$ as random variables.
This follows intuitively from the fact that the posterior variance of a Gaussian process in a fixed location $\vec x_*$ can only decrease with more observations.
\begin{lemma}[The $f_j$s are decreasing in expectation]
\label{lemma:decreasing_expectation}
Assume $\vec x_1, \dots, \vec x_N \in \X$ are independent and identically distributed. 
Denote with $\Proba$ the law of the $\vec x_1,\ldots, \vec x_N$ and with ~$\mat C$ the Cholesky decomposition of $\mat A$.
Define the probability space $(\X, \sigma(\vec x_1, \ldots, \vec x_N), \Proba)$ and the canonical filtration $\mathcal{F}_j\ce \sigma(\vec x_1,\dots, \vec x_j)$ for $j=1, \ldots, N$.
Then the $f_j$ decrease in conditional expectation, that is,  
\[
\Exp[f_{j+1}\mid \sigma(\vec x_1, \ldots, \vec x_{j})]\leq \Exp[f_j\mid \sigma(\vec x_1, \ldots, \vec x_{j-1})]
\, .
\]
\end{lemma}

\newcommand{\vj}[1][j]{\mat C_{#1#1}^2}%
\newcommand{\vjinv}{c} %
\newcommand{\pj}[1][j]{q_{#1}(\vec x)}

\begin{proof}
	Denote with $\mathbb{Q}_j(\dx{\vec x})\ce \proba{\dx{\vec x}\mid \vec x_1, \dots, \vec x_j}$, the regular conditional probability.
	Define the shorthand %
	$\pj\ce\kj\Trans\inv{(\mat K_{j}+\sigma^2\mat I)}\kj$.
	We will show later in the proof, in Eq.~\eqref{eq:pj_decrease}, that $\pj=\pj[j-1]+r_{j}(\vec x)$ where $r_{j}(\vec x)\geq 0$.
	Taking Eq.~\eqref{eq:pj_decrease} as granted for now, we can show the claim as follows.
	\begin{align}%
		\Exp[f_{j+1}\mid \sigma(\vec x_1, \ldots, \vec x_j)]
		&=\Exp[\log \mat C_{j+1,j+1}^2\mid \sigma(\vec x_1, \ldots, \vec x_j)]
		\eqcomment{definition of $f_j$}
		\\&=\int \log\left(k(\vec x, \vec x)+\sigma^2-\kj\Trans(\vec K_j+\sigma^2\vec I)^{-1}\kj\right)\ \mathbb{Q}_{j}(\dx{\vec x}) \eqcomment{property of conditional expectation}
		\\&=\int \log\left(k(\vec x, \vec x)+\sigma^2-\pj\right)\ \mathbb{Q}_{j}(\dx{\vec x}) \eqcomment{definition of $\pj$}
		\\&=\int \log\left(k(\vec x, \vec x)+\sigma^2-\pj[j-1] - r_{j}(\vec x)\right)\ \mathbb{Q}_{j}(\dx{\vec x}) \eqcomment{using Eq.~\eqref{eq:pj_recursion}}
		\\&\leq \int \log\left(k(\vec x, \vec x)+\sigma^2-\pj[j-1]\right)\ \mathbb{Q}_{j}(\dx{\vec x}) \eqcomment{using Eq.~\eqref{eq:pj_decrease} and monotonicity of the logarithm}
		\\&=\int \log\left(k(\vec x, \vec x)+\sigma^2-\pj[j-1]\right)\ \mathbb{Q}_{j-1}(\dx{\vec x}) \eqcomment{with Fubini's theorem}
		\\&= \Exp[\log\vj\mid \sigma(\vec x_1, \ldots, \vec x_{j-1})]
		\eqcomment{property of conditional expectation}
		\\&=\Exp[f_j\mid \sigma(\vec x_1, \ldots, \vec x_{j-1})]
		\eqcomment{definition of $f_j$}
	\end{align}%
	It remains to show $\pj=\pj[j-1]+r_{j}(\vec x)$ where $r_{j}(\vec x)\geq 0$.
	For readability, we define $\vec v_{\vec x}\ce (\vec K_{j-1}+\sigma^2\vec I)^{-1}\kj[j-1]$ and $c\ce v_j^{\!-\!1}$.
	First note, that using block-matrix inversion we can write
	\begin{align}%
		(\vec K_j+\sigma^2\vec I_j)^{-1}=
		\begin{bmatrix}
			(\vec K_{j-1}+\sigma^2\vec I_{j-1})^{-1}+\vec v_{\vec x_j}\vjinv\vec v_{\vec x_j}\Trans & -\vec v_{\vec x_j}\vjinv
			\\-\vec v_{\vec x_j}\Trans \vjinv & \vjinv
		\end{bmatrix}.
	\end{align}%
	Using above observation, we can transform $\pj$.
	\begin{align}%
		\pj&=
		\begin{bmatrix}
			\kj[j-1]\Trans & k(\vec x_j, \vec x)
		\end{bmatrix}
		\\&\quad\cdot
		\begin{bmatrix}
			(\vec K_{j-1}+\sigma^2\vec I)^{-1}+\vec v_{\vec x_j}\vjinv\vec v_{\vec x_j}\Trans & -\vec v_{\vec x_j}\vjinv
			\\-\vec v_{\vec x_j}\Trans \vjinv & \vjinv
		\end{bmatrix}
		\\&\quad\cdot
		\begin{bmatrix}
			\kj[j-1] \\ k(\vec x_j, \vec x)
		\end{bmatrix}
		\eqcomment{definition of $\pj$ and using above observation}
		\\&=
		\begin{bmatrix}
			\kj[j-1]\Trans & k(\vec x, \vec x_j)
		\end{bmatrix}
		\\&\quad\cdot
		\begin{bmatrix}
			\vec v_{\vec x}+\vec v_{\vec x_j}\vjinv\vec v_{\vec x_j}\Trans\kj[j-1] -\vec v_{\vec x_j}\vjinv k(\vec x, \vec x_j)
			\\-\vec v_{\vec x_j}\Trans \kj[j-1] \vjinv +  \vjinv k(\vec x, \vec x_j)
		\end{bmatrix}
		\eqcomment{evaluating the RHS matrix-vector multiplication}
		\\&=
		\kj[j-1]\Trans\vec v_{\vec x}+\vjinv(\vec v_{\vec x_j}\Trans\kj[j-1])^2 
		\\&\quad -2\vec v_{\vec x_j}\Trans\kj[j-1]\vjinv k(\vec x, \vec x_j) + \vjinv k(\vec x, \vec x_j)^2
		\eqcomment{evaluating the vector product}
		\\&=\kj[j-1]\Trans\vec v_{\vec x}+\vjinv(k(\vec x, \vec x_j)-\vec v_{\vec x_j}\Trans \kj[j-1])^2 
		\eqcomment{rearranging terms into a quadratic}
		\\&=\pj[j-1]+\vjinv(k(\vec x, \vec x_j)-\vec v_{\vec x_j}\Trans \kj[j-1])^2
		\eqcomment{definition of $\pj[j-1]$}
	\end{align}%
	This shows that 
	\begin{align}
		\pj&=\pj[j-1]+r_{j}(\vec x) \text{ , where} \label{eq:pj_recursion}\\
		r_{j}(\vec x)&\ce\vjinv(k(\vec x, \vec x_j)-\vec v_{\vec x_j}\Trans \kj[j-1])^2\geq 0. \label{eq:pj_decrease}
	\end{align}
\end{proof}
The claim of Lemma~\ref{lemma:determinant_from_chol} can for example be found in \citet[p.~203]{RasmussenWilliams}.
\begin{lemma}[Computing the log determinant from the Cholesky decomposition]
	\label{lemma:determinant_from_chol}
	Denote with $\mat C$ the Cholesky decomposition of a symmetric and positive definite matrix $\mat A$.
	Then
	\[
	\log\det{\mat A}=2\sum_{j=1}^N\log \mat C_{jj}
	\, .
	\]
\end{lemma}
\begin{proof}
\begin{align}%
	\log|\mat A|&=\log|\mat C\mat C\Trans| \eqcomment{using $\mat K = \mat C\mat C \Trans$}
	\\&=\log(|\mat C|\cdot |\mat C\Trans|) \eqcomment{property of the determinant}
	\\&=\log(|\mat C|^2) \eqcomment{transposition does not affect the determinant}
	\\&=\log \left(\prod_{j=1}^N\mat C_{jj})\right)^2 \eqcomment{property of triangular matrices}
	\\&=2\sum_{j=1}^N\log \mat C_{jj}
	\eqcomment{property of the logarithm}  %
\end{align}%
\end{proof}

\section{Background Material for the Proof of Theorem~\ref{thm:determinant_stopping}}
Before we state the proof of \cref{thm:determinant_stopping}, we provide here the tools that we are going to use.

Our main tool will be the following theorem by \citet{Fan2012Hoeffding}.
Essentially, this is a generalization of Hoeffding's inequality to martingales.
It states that for a sum of random variables that decrease in (conditional) expectation, the probability of exceeding a certain threshold is low, when at the same time the (conditional) variance is bounded by another constant.
Importantly, this probability holds \emph{simultaneously} for all partial sums starting in~$1$ and ending in $n=1$ to $n=N$.
\begin{theorem}[Hoeffding's inequality for supermartingales \citep{Fan2012Hoeffding}]\label{theorem:fan}\ \\	
	As\-sume that $(\xi_j, \mathcal{F}_j)_{j=1, \ldots, N}$ are supermartingale differences satisfying $\xi_j\leq 1$.
	Then, for any $x\geq 0$ and $v > 0$, 
	\begin{align}%
		&\mathbb{P}\Big(\text{ for some } n \in [1, N]
		\\&\qquad \sum_{j=1}^n \xi_j \geq x \text{ and } \sum_{j=1}^n \Var[\xi_j \mid \mathcal{F}_{j-1}]\leq v \Big)\leq H_N(x, v),
	\end{align}%
	where
	\[
	H_N(x, v)\ce \one_{\{x\leq N\}} \left\{\left(\frac{v}{v+x}\right)^{v+x}\left(\frac{N}{N-x}\right)^{N-x}\right\}^{\frac{N}{N+v}}
	\, .
	\]
\end{theorem}

\label{app:lemmata}
The following theorem will give us the upper bound on the conditional variance, necessary for Theorem~\ref{theorem:fan}.
Below theorem applies to empirical variance estimates, but the remark below shows that this is also a bound on the true variance.
\begin{theorem}[Popoviciu's inequality {\citep{Popoviciu1935variance, Sharma2010variance}}]
	\label{thm:popoviciu}
	For a sequence of real numbers $x_1, ..., x_n\in [m, M]$, define
	$\mu\ce \frac{1}{n}\sum_{j=1}^n x_j$ and $\sigma^2\ce \frac{1}{n}\sum_{j=1}^N (x_j-\mu)^2$, then
	$$\sigma^2\leq \nicefrac{1}{4}(M-m)^2.$$
\end{theorem}
\begin{remark}
	\cref{thm:popoviciu} can be used to obtain a bound on the conditional variance as well.
	Let $x_1,...,x_n\sim P(\cdot \mid\mathcal{F})$ be independent.
	Then,
	\begin{align}%
		\Var[X\mid\mathcal{F}]&=\Exp[(X-\Exp[X\mid\mathcal{F}])^2\mid\mathcal{F}]
		\eqcomment{definition of conditional variance}
		\\&=\frac{n}{n-1}\Exp[\sigma^2\mid\mathcal{F}] \eqcomment{using Bessel's correction}
		\\&\leq \frac{n}{4(n-1)}(M-m)^2
		\eqcomment{by Theorem~\ref{thm:popoviciu}}
	\end{align}%
	which holds for all $n \in \mathbb{N}$.
	Hence, $\Var [X\mid\mathcal{F}]\leq \nicefrac{1}{4}(M-m)^2$.
\end{remark}
The martingale differences that we will be analyzing have random indices from our stopping time.
Doob's Optional Sampling Theorem (see for example \textcite[p.~489]{Grimmett2001RandomProcesses}) and the remark below provide us with the mathematical justification.
\begin{theorem}[{Doob's Optional Sampling Theorem}] 
	\label{thm:optional_sampling}
	Let $(X_j, \mathcal{F}_j)_{j\in \mathbb{N}}$ be a submartingale and $\tau_1\leq\tau_2\leq ...$ be a sequence of stopping times s.t.~$P(\tau_j\leq n_j)=1$ for some deterministic real sequence $n_j$, then the stopped process $(X_{\tau_j}, \mathcal{F}_{\tau_j})_{j\in\mathbb{N}}$ is also a submartingale.
\end{theorem}
\begin{remark}
	By exchanging $X_j$ for $-X_j$ the theorem can be shown to hold for supermartingales as well.
\end{remark}
\begin{corollary}[Stopped submartingale differences]
	\label{corr:stopped_martingale_difference}
	Let $(\xi_j, \mathcal{F}_j)_{j\in \mathbb{N}}$ be a submartingale-difference and let $\tau$ be a stopping time, then the stopped process $(\xi_{\min(j,\tau)}, \mathcal{F}_{\min(j,\tau)})_{j\in\mathbb{N}}$ is also a submartingale-difference.
\end{corollary}
\begin{proof}
	Define $X_l\ce \sum_{j=1}^l \xi_j$ and observe that this defines a submartingale.
	By Theorem \ref{thm:optional_sampling} $(X_{\min(j,\tau)},\mathcal{F}_{\min(j,\tau)})_{j\in\mathbb{N}}$ is a submartingale.
	Then $X_{\min(j,\tau)}-X_{\min(j,\tau)-1}=\xi_{\min(j,\tau)}$ is again a submartingale-difference.
\end{proof}

\section{Proof of Theorem~\ref{thm:determinant_stopping}}
\label{app:proof_main_thm}
The proof can be split into two parts.
Lemma~\ref{lemma:when_bound_holds} shows by using the stopping conditions that if the bound holds, the relative error of the estimator is indeed less than~$r$ with probability~1.
The second part is to show that $\proba{\U_\tau < D_N}\leq \delta$, which is the purpose of Lemma~\ref{lemma:bound_probably_holds} and which makes use of the assumption stated in Eq.~\eqref{eq:det_crucial_assumption}. 

\begin{proof}
\begin{align}
	&\proba{\abs{\frac{D_N-\hat{D}_\tau}{D_N}}>r} 
	\\ &=\proba{\frac{\smallabs{D_N-\hat{D}_\tau}}{\abs{D_N}}>r, D_N \leq \U_\tau } \notag
	\\&\quad + \proba{\frac{\smallabs{D_N-\hat{D}_\tau}}{\abs{D_N}}>r, D_N > \U_\tau}  \eqcomment{sum rule}
	\\&\leq \proba{\frac{\smallabs{D_N-\hat{D}_\tau}}{\abs{D_N}}>r, D_N \leq \U_\tau}
	+ \proba{D_N > \U_\tau} \label{eq:main} \eqcomment{upper-bounding joint by marginal}
	\\&\leq 0 + \delta
	\eqcomment{by Lemma~\ref{lemma:when_bound_holds} and Lemma~\ref{lemma:bound_probably_holds}} %
\end{align}
\end{proof}
The following lemma gives an upper bound on the relative error of an estimator in terms of upper and lower bounds for the quantity of interest.
The bound is minimized if the estimator is chosen to be the average of upper and lower bound.
The lemma can also be found in \citet{Mnih2008EBSmasterThesis} but has been developed independently.
\begin{lemma}[Bounding the relative error]
	\label{lemma:rel_err_bound}
	Let $D, \hat{D}\in [\L, \U]$, and assume $\sign(\L)=\sign(\U)\neq 0$. 
	Then the relative error of the estimator $\hat{D}$ can be bounded as 
	\[
	\frac{\smallabs{D-\hat{D}}}{\abs{D}}\leq \frac{\max(\U-\hat{D}, \hat{D}-\L)}{\min(\abs{\L}, \abs{\U})}
	\, .
	\]
\end{lemma}
\begin{proof}
	First observe that if $D_N>\hat{D}$ then $\smallabs{D_N-\hat{D}}=D_N-\hat{D}\leq \U-\hat{D}$.
	If $D_N\leq \hat{D}$, then $\smallabs{D_N-\hat{D}}=\hat{D}-D_N\leq \hat{D}-\L$.
	Hence, $$\smallabs{D_N-\hat{D}}\leq \max(\U-\hat{D}, \hat{D}-\L).$$
	Case $\L>0$: In this case $\smallabs{D_N}=D_N\geq \L=\smallabs{\L}$, and we obtain for the relative error:
	\begin{align}%
		\frac{\max(\U-\hat{D}, \hat{D}-\L)}{\smallabs{D_N}}&\leq \frac{\max(\U-\hat{D}, \hat{D}-\L)}{\smallabs{\L}}\, .
	\end{align}%
	
	Case $\U<0$: In that case $\smallabs{\L}\geq \smallabs{D_N}\geq \smallabs{\U}$, and the relative error can be bounded as follows.
	\begin{align}%
		\frac{\max(\U-\hat{D}, \hat{D}-\L)}{\smallabs{D_N}}&\leq\frac{\max(\U-\hat{D}, \hat{D}-\L)}{\smallabs{\U}}
	\end{align}%
	Since we assumed $\sign(\L)=\sign(\U)$ these were all cases that required consideration.
	Combining all observations yields
	\begin{align}%
		\frac{\smallabs{D_N-\hat{D}}}{\smallabs{D_N}}
		&\leq \max(\U-\hat{D}, \hat{D}-\L)\max\left(\frac{1}{\smallabs{\U}}, \frac{1}{\smallabs{\L}}\right)
		\\&= \frac{\max(\U-\hat{D}, \hat{D}-\L)}{\min(\smallabs{\U}, \smallabs{\L})}\ .
	\end{align}%
\end{proof}

\begin{lemma}[Controlling the relative error when $D_N \leq \U_\tau$]
	\label{lemma:when_bound_holds}
	With the definitions of Section~\ref{sec:empirical_estimator_problem_definition}, the probability that the relative error of the estimator is larger than some $r>0$ and at the same time the bound holds, is zero.
	Formally,
	\[
	\proba{\frac{\smallabs{D_N-\hat{D}_\tau}}{\smallabs{D_N}}>r, D_N \leq \U_\tau }=0\, .
	\]
\end{lemma}

\begin{proof}
As a preliminary observation note that
\begin{align}
	D_N &= \sum_{j=1}^N f_{j} \notag 
	\eqcomment{by definition} 
	\\&=D_n + \sum_{j=n+1}^N f_j \text{ for all $n=0, \ldots, N$} \notag
	\eqcomment{definition of $D_n$}
	\\&\geq D_n+\sum_{j=n+1}^N \lBound^- \text{ for all $n=0, \ldots, N$} \notag
	\eqcomment{since $f_j\in[\lBound^-, \lBound^+]$}
	\\
	\label{eq:det_as_lower_bound}
	&=\L_n \text{ for all $n=0, \ldots, N$} 
	\eqcomment{using the definition of $\L_n$}
\end{align}
and hence, for all $n=0, \ldots, N$, $\L_n$ is an almost sure lower bound to $D_N$.
\begin{align}%
	&\proba{\frac{\smallabs{D_N-\hat{D}_\tau}}{\abs{D_N}}>r, D_N \leq \U_\tau }
	\\&=\quad\proba{\frac{\smallabs{D_N-\hat{D}_\tau}}{\abs{D_N}}>r, D_N \leq \U_\tau, \tau<N}
	\\&\quad+\proba{\frac{\smallabs{D_N-\hat{D}_\tau}}{\abs{D_N}}>r, D_N \leq \U_\tau, \tau =N}
\end{align}%
Recall that $\hat{D}_\tau= \nicefrac{1}{2}(\L_\tau + \U_\tau)$.
In case $\tau=N$, we have that $\U_N=\L_N=D_N$, and hence, $\proba{\frac{\smallabs{D_N-\hat{D}_\tau}}{\abs{D_N}}>r, D_N \leq \U_\tau, \tau =N}=0$.

\newcommand{\evA}{A}%
For brevity, define the event $\evA\ce \{D_N \leq \U_\tau, \tau < N\}$, that is, the upper bound holds and the stopping conditions are fulfilled at a time before $N$.

\begin{align}%
	&\proba{\frac{\smallabs{D_N-\hat{D}_\tau}}{\abs{D_N}}>r, D_N \leq \U_\tau, \tau<N}
	\\&= \proba{\frac{\smallabs{D_N-\hat{D}_\tau}}{\abs{D_N}}>r, \evA{}}
	\eqcomment{definition of $\evA{}$}
	\\& = \proba{\frac{\smallabs{D_N-\hat{D}_\tau}}{\abs{D_N}}>r, \L_\tau\leq D_N, \evA{}} 
	\eqcomment{since $\L_\tau$ is an almost sure lower bound to $D_N$ by Eq.~\eqref{eq:det_as_lower_bound}}
	\\&\leq \proba{\frac{\max(\U_\tau-\hat{D}_\tau, \hat{D}_\tau-\L_\tau)}{\min(\abs{\L_\tau},\abs{\U_\tau})}>r, \L_\tau\leq D_N, \evA{}}
	\eqcomment{by Lemma~\ref{lemma:rel_err_bound}, using the first condition of $\tau$, Eq.~\eqref{eq:condition_sign}}
	\\&= \proba{\frac{\U_\tau-\L_\tau}{2\min(\abs{\L_\tau},\abs{\U_\tau})}>r, \L_\tau\leq D_N, \evA{}}
	\eqcomment{definition of $\hat{D}_\tau$}
	\\&=0 \eqcomment{by the second condition of $\tau$, Eq.~\eqref{eq:condition_relative}} %
\end{align}%
\end{proof}

\begin{lemma}[Upper bound control]
	\label{lemma:bound_probably_holds}
	With the definitions of Section~\ref{sec:empirical_estimator_problem_definition}, the probability that the upper bound fails is less than $\delta$.
	Formally, 
	$$\proba{D_N > \U_\tau} \leq \delta.$$
\end{lemma}
\begin{proof}
The following parts of the proof rely on Theorem~\ref{theorem:fan} by \citet{Fan2012Hoeffding}.
To apply Theorem~\ref{theorem:fan}, define $Z_j'\ce f_{j}-\Exp[f_{j}\mid \mathcal{F}_{j-1}]$ and $Z_j\ce Z'_{\tau+j}$.

For brevity, we define $\epsilon\ce (\lBound^+-\lBound^-) H_N^{-1}(\nicefrac{\delta}{2})$, $\epsilon_n\ce\epsilon\left(\frac{1}{N-n}+\frac{1}{n}\right)$, and $\me[n]\ce \frac{D_n}{n}+\epsilon_n$ such that we can write 
\begin{align}
	\label{eq:convenient_U}
	\U_n= D_n + (N-n)\min(\me[n], \lBound^+)\, .
\end{align}
\begin{align}
	& \proba{D_N>\U_\tau} \notag
	\\*=& \proba{D_\tau+\sum_{j=\tau+1}^N \l{j}>D_\tau+(N-\tau)\min\left(\me, \lBound^+\right)} \notag
	\eqcomment{using the definition of $D_n$ and Eq.~\eqref{eq:convenient_U}}
	\\=& \proba{\sum_{j=\tau+1}^N \l{j}>(N-\tau)\min\left(\me, \lBound^+\right)} \notag
	\eqcomment{simplifying}
	\\=& \proba{\sum_{j=\tau+1}^N \l{j}>(N-\tau)\me \text{ or } \sum_{j=\tau+1}^N \l{j}>(N-\tau)\lBound^+} \notag
	\eqcomment{exchanging $\min$ for logical or}
	\\= & \proba{\sum_{j=\tau+1}^N \l{j}>(N-\tau)\me} \notag
	\eqcomment{since $\l{j}\leq \lBound^+$}
	\\= & \proba{\sum_{j=1}^{N-\tau} \left[Z_{j}+\Exp[\l{\tau+j}\mid\mathcal{F}_{\tau+j-1}]\right]>(N-\tau)\me} \notag
	\eqcomment{definition of $Z_j$}
	\\ 
	\leq &\Proba\left(\sum_{j=1}^{N-\tau} Z_{j}+\sum_{j=\tau+1}^{N}\Exp[\l{j}\mid\mathcal{F}_{j-1}]>(N-\tau)\me, \right. \notag
	\\&\quad\left.\sum_{j=\tau+1}^{N}\Exp[\l{j}\mid\mathcal{F}_{j-1}]\leq\frac{N-\tau}{\tau}\left({D_\tau+\epsilon}\right)\right)\notag
	\\&+\proba{\sum_{j=\tau+1}^{N}\Exp[\l{j}\mid\mathcal{F}_{j-1}]>\frac{N-\tau}{\tau}\left({D_\tau+\epsilon}\right)} 
	\label{eq:azuma_problem}
	\eqcomment{sum rule and upper-bounding joint by marginal}
\end{align}
Consider the first addend in Eq.~\eqref{eq:azuma_problem}.
\begin{align}
	& \Proba\left(\sum_{j=1}^{N-\tau} Z_{j}+\sum_{j=\tau+1}^{N}\Exp[\l{j}\mid\mathcal{F}_{j-1}]>(N-\tau)\me, \right. \notag
	\\&\quad\left.\sum_{j=\tau+1}^{N}\Exp[\l{j}\mid\mathcal{F}_{j-1}]\leq\frac{N-\tau}{\tau}\left({D_\tau+\epsilon}\right)\right)\notag
	\\&\leq \proba{\sum_{j=1}^{N-\tau} Z_{j}+\frac{N-\tau}{\tau}\left({D_\tau+\epsilon}\right)>(N-\tau)\me}\notag \eqcomment{combining the two events}
	\\&= \mathbb{P}\left(\sum_{j=1}^{N-\tau} Z_{j}+\frac{N-\tau}{\tau}\left({D_\tau+\epsilon}\right)> \right. \notag
	\\*&\qquad\qquad\qquad(N-\tau)\left(\frac{D_\tau}{\tau}+\epsilon\left(\frac{1}{N-\tau}+\frac{1}{\tau}\right)\right)\Bigg)\notag
	\eqcomment{definition of $\me$ and $\epsilon_n$}
	\\&= \proba{\sum_{j=1}^{N-\tau} Z_{j}>\epsilon} \notag \eqcomment{simplifying}
	\\&= \proba{\sum_{j=1}^{N-\tau} \frac{Z_{j}}{\lBound^+-\lBound^-}> H_N^{-1}(\nicefrac{\delta}{2})} \label{eq:fan_application}
	\eqcomment{definition of $\epsilon$ and dividing by $\lBound^+-\lBound^-$}
	\\&\leq \mathbb{P}\left( \sum_{j=1}^{n} \frac{Z_{j}}{\lBound^+-\lBound^-}>H_N^{-1}(\nicefrac{\delta}{2})\right.\notag
	\\*&\qquad\qquad\qquad\text{ for some } n\in \{1, \ldots, N\}\Bigg) \notag
	\eqcomment{enlarging the event}
\end{align}
We are now ready to use Theorem~\ref{theorem:fan}.
Since $(Z'_j, \mathcal{F}_j)_{j\in \{1, \dots, N\}}$ is a martingale difference, 
\[
\left(Z_{\min(j, N)}, \mathcal{F}_{\min(\tau+j,N)}\right)_{j\in \mathbb{N}_0}
\]
is a martingale difference as well (Corollary~\ref{corr:stopped_martingale_difference}).
Further note, that the random variables $\frac{Z_j}{\lBound^+-\lBound^-}$ are bounded from above by $1$.
Hence, there is only one ingredient missing to apply Theorem~\ref{theorem:fan}, which is a bound on the conditional variance.
To this end, we use Popoviciu's inequality.
The latter is applicable, since the $\frac{Z_j}{\lBound^+-\lBound^-}$ are also bounded from below by $-1$.
\begin{align}
	&\proba{\sum_{j=1}^{n} \frac{Z_{j}}{\lBound^+-\lBound^-}>H_N^{-1}(\nicefrac{\delta}{2}) \text{ for some } n\in \{1, \ldots, N\}} \notag
	\\&=\mathbb{P}\left(\sum_{j=1}^{n} \frac{Z_{j}}{\lBound^+-\lBound^-}>H_N^{-1}(\nicefrac{\delta}{2}), \right.\notag
	\\*&\qquad\quad\left.\sum_{j=1}^n \Var\left[\left.\frac{Z_j}{\lBound^+-\lBound^-}\right\rvert \mathcal{F}_{j-1}\right]\leq N \right. \notag
	\\*&\qquad\qquad\qquad \text{ for some } n\in \{1, \ldots, N\}\Bigg) \notag \eqcomment{by Popoviciu's inequality (Theorem~\ref{thm:popoviciu})}
	\\&\leq H(H_N^{-1}(\nicefrac{\delta}{2}), N) \notag
	\eqcomment{by Theorem \ref{theorem:fan}, where $H$ is defined in that theorem}
	\\ &= H_N(H_N^{-1}(\nicefrac{\delta}{2}))\leq \nicefrac{\delta}{2}. \notag
	\eqcomment{definition of $H_N$}
\end{align}

Now we will take care of the second addend in Eq.~\eqref{eq:azuma_problem}, using the assumption that the $\l{j}$ decrease in expectation: Eq.~\eqref{eq:det_crucial_assumption}.
We will again apply Theorem \ref{theorem:fan}.
\begin{align}%
	&\proba{\sum_{j=\tau+1}^{N}\Exp[\l{j}\mid\mathcal{F}_{j-1}]>\frac{N-\tau}{\tau}\left({D_\tau+\epsilon}\right)} 
	\\&\leq \proba{\sum_{j=\tau+1}^{N}\Exp[\l{\tau+1}\mid\mathcal{F}_{\tau}]>\frac{N-\tau}{\tau}\left({D_\tau+\epsilon}\right)} \eqcomment{using Eq.~\eqref{eq:det_crucial_assumption}}
	\\&= \proba{\tau \Exp[l_{\tau+1}\mid\mathcal{F}_{\tau}]> D_\tau+\epsilon} 
	\eqcomment{dividing by $N-\tau$ and multiplying by $\tau$}
	\\&= \proba{\sum_{j=1}^\tau(\Exp[l_{\tau+1}\mid\mathcal{F}_{\tau}]-\l{j})>\epsilon}
	\eqcomment{definition of $D_\tau$}
	\\&\leq \proba{\sum_{j=1}^\tau(\Exp[l_{j+1}\mid\mathcal{F}_{j}]-\l{j})>\epsilon}
	\eqcomment{using again  Eq.~\eqref{eq:det_crucial_assumption}}
	\\&=\proba{\sum_{j=1}^\tau \frac{\Exp[l_{j+1}\mid\mathcal{F}_{j}]-\l{j}}{\lBound^+-\lBound^-}>H_N^{-1}\left(\frac{\delta}{2}\right)}
	\eqcomment{definition of $\epsilon$ and dividing by $\lBound^+-\lBound^-$}
	\\&= \proba{\sum_{j=1}^\tau-\frac{Z_j'}{\lBound^+-\lBound^-}> H_N^{-1}\left(\frac{\delta}{2}\right)} \eqcomment{definition of $Z_j'$} 
\end{align}%
Changing the sign does not change the martingale difference property and hence, $(-Z_j', \mathcal{F}_j)_{j\in \{1, \dots, N\}}$ is a martingale difference as well.
We can apply the same argument as in Eq.~\eqref{eq:fan_application}.
\begin{align}%
	&\proba{\sum_{j=1}^{n}-\frac{Z_{j}'}{\lBound^+-\lBound^-}> H_N^{-1}\left(\frac{\delta}{2}\right)}
	\\&\leq H\left(H_N^{-1}\left(\frac{\delta}{2}\right), N\right) \eqcomment{using the same argument as in Eq.~\eqref{eq:fan_application}}
	\\&
	\leq \frac{\delta}{2} %
\end{align}%
\end{proof}
\section{Results}
\label{app:results}
This section contains the complete results from the experiments described in \cref{sec:experiments} in  \cref{fig:results_time_protein,fig:results_time_tamilnadu_electricity,fig:results_time_bank,fig:results_time_metro,fig:results_time_pumadyn}.
Considering the same datasets, \cref{fig:ste} shows the relative error when using the method of \citet{Gardner2018gpytorch} with default parameters for the RBF kernel.

\begin{figure}%
	{\centering\ref{leg:performance}\\}
	\setlength{\tabcolsep}{-4pt}
	\begin{tabular}{rr}
		\multicolumn{1}{c}{RBF \cref{eq:kernel_rbf}}&\multicolumn{1}{c}{OU \cref{eq:kernel_ou}}\\
		\renewcommand{\labely}{\metric}%
		\input{tikz/RBF/protein/joint_plot_-1.0.tikz}
		&
		\input{tikz/OU/protein/joint_plot_-1.0.tikz}\\
		\renewcommand{\labely}{\metric}%
		\input{tikz/RBF/protein/joint_plot_0.0.tikz}
		&
		\input{tikz/OU/protein/joint_plot_0.0.tikz}\\
		\renewcommand{\labely}{\metric}%
		\input{tikz/RBF/protein/joint_plot_1.0.tikz}
		&
		\input{tikz/OU/protein/joint_plot_1.0.tikz}\\
		\renewcommand{\labely}{\metric}%
		\input{tikz/RBF/protein/joint_plot_2.0.tikz}
		&
		\input{tikz/OU/protein/joint_plot_2.0.tikz}
	\end{tabular}
	\caption{
		\captionText{\PROTEIN}
	}
	\label{fig:results_time_protein}
\end{figure}

\begin{figure}%
		{\centering\ref{leg:performance}\\}
	\setlength{\tabcolsep}{-4pt}
	\begin{tabular}{rr}
		\multicolumn{1}{c}{RBF \cref{eq:kernel_rbf}}&\multicolumn{1}{c}{OU \cref{eq:kernel_ou}}\\
		\renewcommand{\labely}{\metric}%
		\input{tikz/RBF/tamilnadu_electricity/joint_plot_-1.0.tikz}
		&
		\input{tikz/OU/tamilnadu_electricity/joint_plot_-1.0.tikz}\\
		\renewcommand{\labely}{\metric}%
		\input{tikz/RBF/tamilnadu_electricity/joint_plot_0.0.tikz}
		&
		\input{tikz/OU/tamilnadu_electricity/joint_plot_0.0.tikz}\\
		\renewcommand{\labely}{\metric}%
		\input{tikz/RBF/tamilnadu_electricity/joint_plot_1.0.tikz}
		&
		\input{tikz/OU/tamilnadu_electricity/joint_plot_1.0.tikz}\\
		\renewcommand{\labely}{\metric}%
		\input{tikz/RBF/tamilnadu_electricity/joint_plot_2.0.tikz}
		&
		\input{tikz/OU/tamilnadu_electricity/joint_plot_2.0.tikz}
	\end{tabular}
	\caption{
		\captionText{\TAMILNADU}
	}
	\label{fig:results_time_tamilnadu_electricity}
\end{figure}

\begin{figure}%
		{\centering\ref{leg:performance}\\}
	\setlength{\tabcolsep}{-4pt}
	\begin{tabular}{rr}
		\multicolumn{1}{c}{RBF \cref{eq:kernel_rbf}}&\multicolumn{1}{c}{OU \cref{eq:kernel_ou}}\\
		\renewcommand{\labely}{\metric}%
		\input{tikz/RBF/bank/joint_plot_-1.0.tikz}
		&
		\input{tikz/OU/bank/joint_plot_-1.0.tikz}\\
		\renewcommand{\labely}{\metric}%
		\input{tikz/RBF/bank/joint_plot_0.0.tikz}
		&
		\input{tikz/OU/bank/joint_plot_0.0.tikz}\\
		\renewcommand{\labely}{\metric}%
		\input{tikz/RBF/bank/joint_plot_1.0.tikz}
		&
		\input{tikz/OU/bank/joint_plot_1.0.tikz}\\
		\renewcommand{\labely}{\metric}%
		\input{tikz/RBF/bank/joint_plot_2.0.tikz}
		&
		\input{tikz/OU/bank/joint_plot_2.0.tikz}
	\end{tabular}
	\caption{
		\captionText{\BANK}
	}
	\label{fig:results_time_bank}
\end{figure}

\begin{figure}%
		{\centering\ref{leg:performance}\\}
	\setlength{\tabcolsep}{-4pt}
	\begin{tabular}{rr}
		\multicolumn{1}{c}{RBF \cref{eq:kernel_rbf}}&\multicolumn{1}{c}{OU \cref{eq:kernel_ou}}\\
		\renewcommand{\labely}{\metric}%
		\input{tikz/RBF/metro/joint_plot_-1.0.tikz}
		&
		\input{tikz/OU/metro/joint_plot_-1.0.tikz}\\
		\renewcommand{\labely}{\metric}%
		\input{tikz/RBF/metro/joint_plot_0.0.tikz}
		&
		\input{tikz/OU/metro/joint_plot_0.0.tikz}\\
		\renewcommand{\labely}{\metric}%
		\input{tikz/RBF/metro/joint_plot_1.0.tikz}
		&
		\input{tikz/OU/metro/joint_plot_1.0.tikz}\\
		\renewcommand{\labely}{\metric}%
		\input{tikz/RBF/metro/joint_plot_2.0.tikz}
		&
		\input{tikz/OU/metro/joint_plot_2.0.tikz}
	\end{tabular}
	\caption{
		\captionText{\METRO}
	}
	\label{fig:results_time_metro}
\end{figure}

\begin{figure}%
		{\centering\ref{leg:performance}\\}
	\setlength{\tabcolsep}{-4pt}
	\begin{tabular}{rr}
		\multicolumn{1}{c}{RBF \cref{eq:kernel_rbf}}&\multicolumn{1}{c}{OU \cref{eq:kernel_ou}}\\
		\renewcommand{\labely}{\metric}%
		\input{tikz/RBF/pumadyn/joint_plot_-1.0.tikz}
		&
		\input{tikz/OU/pumadyn/joint_plot_-1.0.tikz}\\
		\renewcommand{\labely}{\metric}%
		\input{tikz/RBF/pumadyn/joint_plot_0.0.tikz}
		&
		\input{tikz/OU/pumadyn/joint_plot_0.0.tikz}\\
		\renewcommand{\labely}{\metric}%
		\input{tikz/RBF/pumadyn/joint_plot_1.0.tikz}
		&
		\input{tikz/OU/pumadyn/joint_plot_1.0.tikz}\\
		\renewcommand{\labely}{\metric}%
		\input{tikz/RBF/pumadyn/joint_plot_2.0.tikz}
		&
		\input{tikz/OU/pumadyn/joint_plot_2.0.tikz}
	\end{tabular}
	\caption{
		\captionText{\PUMADYN}
	}
	\label{fig:results_time_pumadyn}
\end{figure}

\begin{figure}
	\renewcommand{\arraystretch}{0.0}%
	\setlength{\figwidth}{.5\textwidth}%
	\setlength{\figheight}{.15\textheight}%
	\renewcommand{\expconfig}[3]{#3}%
	\renewcommand{\ticklabelsx}{false}%
	\renewcommand{\labelx}{~}%
	\begin{tabular}{ll}
		\renewcommand{\ticklabelsy}{true}%
		\renewcommand{\labely}{$r$}%
		\input{tikz/ste_bank_rbf.tikz}%
		&
		\renewcommand{\ticklabelsy}{false}%
		\renewcommand{\labely}{~}%
		\input{tikz/ste_metro_rbf.tikz}%
		\\[-10pt]
		\renewcommand{\ticklabelsy}{true}%
		\renewcommand{\labely}{$r$}%
		\input{tikz/ste_pm25_rbf.tikz}%
		&
		\renewcommand{\ticklabelsy}{false}%
		\renewcommand{\labely}{~}%
		\input{tikz/ste_protein_rbf.tikz}%
		\\[-10pt]
		\renewcommand{\ticklabelsy}{true}%
		\renewcommand{\labely}{$r$}%
		\renewcommand{\ticklabelsx}{true}%
		\renewcommand{\labelx}{$\log(\ell)$}%
		\input{tikz/ste_tamilnadu_electricity_rbf.tikz}%
		&
		\renewcommand{\ticklabelsy}{false}%
		\renewcommand{\labely}{~}%
		\renewcommand{\ticklabelsx}{true}%
		\renewcommand{\labelx}{$\log(\ell)$}%
		\input{tikz/ste_pumadyn_rbf.tikz}%
	\end{tabular}
	\caption{the need for theoretical guarantees. 
		Of the related work described in \cref{sec:related_work} only \citet{Gardner2018gpytorch} provide publicly accessible code.
		The figure shows the achieved relative error $r$, \cref{eq:relative_precision}, when using default parameters, for the RBF kernel, \cref{eq:kernel_rbf}, with $\theta\ce1$ and different length-scales $\ell$ on all our considered datasets (see \cref{tbl:datasets}).
		The relative error is \emph{more often than not} worse than $0.1$ and can differ over two orders of magnitude.
		Theorem 2 in \citet{Gardner2018gpytorch} which could describe how to set the parameters of their method to achieve a desired precision is not applicable in this setting (see \cref{sec:related_work}).
	}
	\label{fig:ste}
\end{figure}

\FloatBarrier
\bibliography{bibfile}

\begin{thebibliography}{33}
\providecommand{\natexlab}[1]{#1}
\providecommand{\url}[1]{\texttt{#1}}
\expandafter\ifx\csname urlstyle\endcsname\relax
  \providecommand{\doi}[1]{doi: #1}\else
  \providecommand{\doi}{doi: \begingroup \urlstyle{rm}\Url}\fi

\bibitem[Boucheron et~al.(2013)Boucheron, Lugosi, and
  Massart]{Boucheron2013concentration}
St\'ephane Boucheron, G\'abor Lugosi, and Pascal Massart.
\newblock \emph{Concentration Inequalities: A Nonasymptotic Theory of
  Independence}.
\newblock Oxford University Press, 1st edition, 2013.

\bibitem[Boutsidis et~al.(2017)Boutsidis, Drineas, Kambadur, Kontopoulou, and
  Zouzias]{Boutsidis2017Determinant}
Christos Boutsidis, Petros Drineas, Prabhanjan Kambadur, Eugenia-Maria
  Kontopoulou, and Anastasios Zouzias.
\newblock A randomized algorithm for approximating the log determinant of a
  symmetric positive definite matrix.
\newblock \emph{Linear Algebra and its Applications}, 533:\penalty0 95 -- 117,
  2017.

\bibitem[Chalupka et~al.(2013)Chalupka, {Williams,~C.~K.~I.}, and
  Murray]{Chalupka2013comparison}
Krzysztof Chalupka, {Williams,~C.~K.~I.}, and Iain Murray.
\newblock A framework for evaluating approximation methods for {G}aussian
  process regression.
\newblock \emph{Journal of Machine Learning Research}, 14\penalty0
  (1):\penalty0 333--350, 2013.

\bibitem[Davidson(1994)]{Davidson1994StochasticLimitTheory}
James Davidson.
\newblock \emph{Stochastic Limit Theory: An Introduction for Econometricians}.
\newblock Oxford University Press, 1994.

\bibitem[Diaconis(1988)]{diaconis88:_bayes}
P.~Diaconis.
\newblock {{B}ayesian numerical analysis}.
\newblock \emph{Statistical decision theory and related topics}, IV\penalty0
  (1):\penalty0 163--175, 1988.

\bibitem[Dong et~al.(2017)Dong, Eriksson, Nickisch, Bindel, and
  Wilson]{Dong2017scalable}
Kun Dong, David Eriksson, Hannes Nickisch, David Bindel, and Andrew~G. Wilson.
\newblock Scalable log determinants for gaussian process kernel learning.
\newblock In \emph{Advances in Neural Information Processing Systems}, pages
  6330--6340, 2017.

\bibitem[Dorn and En{\ss{}}lin(2015)]{Dorn2015Determinant}
Sebastian Dorn and Torsten~A. En{\ss{}}lin.
\newblock Stochastic determination of matrix determinants.
\newblock \emph{Physical Review E}, 92:\penalty0 013302, 2015.

\bibitem[Dua and Graff(2019)]{Dua2019uci}
Dheeru Dua and Casey Graff.
\newblock {UCI} machine learning repository, 2019.
\newblock URL \url{http://archive.ics.uci.edu/ml}.

\bibitem[Fan et~al.(2012)Fan, Grama, and Liu]{Fan2012Hoeffding}
Xiequan Fan, Ion Grama, and Quansheng Liu.
\newblock Hoeffding’s inequality for supermartingales.
\newblock \emph{Stochastic Processes and their Applications}, 122\penalty0
  (10):\penalty0 3545--3559, 2012.

\bibitem[Fitzsimons et~al.(2017{\natexlab{a}})Fitzsimons, Cutajar, Osborne,
  Roberts, and Filippone]{Fitzsimons2017BayesianDeterminant}
Jack Fitzsimons, Kurt Cutajar, Michael Osborne, Stephen Roberts, and Maurizio
  Filippone.
\newblock {B}ayesian inference of log determinants.
\newblock In Gal Elidan, Kristian Kersting, and Alexander~T. Ihler, editors,
  \emph{Thirty-Third Conference on Uncertainty in Artificial Intelligence,
  {UAI} 2017, August 11-15, 2017, Sydney, Australia}, 2017{\natexlab{a}}.

\bibitem[Fitzsimons et~al.(2017{\natexlab{b}})Fitzsimons, Granziol, Cutajar,
  Osborne, Filippone, and Roberts]{Fitzsimons2017EntropicDeterminant}
Jack Fitzsimons, Diego Granziol, Kurt Cutajar, Michael Osborne, Maurizio
  Filippone, and Stephen Roberts.
\newblock Entropic trace estimates for log determinants.
\newblock In Michelangelo Ceci, Jaakko Hollm{\'e}n, Ljup{\v{c}}o Todorovski,
  Celine Vens, and Sa{\v{s}}o D{\v{z}}eroski, editors, \emph{Machine Learning
  and Knowledge Discovery in Databases}, pages 323--338, 2017{\natexlab{b}}.

\bibitem[Gardner et~al.(2018)Gardner, Pleiss, Bindel, Weinberger, and
  Wilson]{Gardner2018gpytorch}
Jacob~R. Gardner, Geoff Pleiss, David Bindel, Kilian~Q. Weinberger, and
  Andrew~G. Wilson.
\newblock Gpytorch: Blackbox matrix-matrix gaussian process inference with gpu
  acceleration.
\newblock In \emph{Advances in Neural Information Processing Systems}, 2018.

\bibitem[George et~al.(1986)George, Heath, and Liu]{George1986parallelCholesky}
Alan George, Michael~T. Heath, and Joseph Liu.
\newblock Parallel cholesky factorization on a shared-memory multiprocessor.
\newblock \emph{Linear Algebra and its Applications}, 77:\penalty0 165--187,
  1986.

\bibitem[Golub and {Van Loan}(2013)]{golub2013matrix4}
Gene~H. Golub and Charles~F. {Van Loan}.
\newblock \emph{{Matrix computations}}.
\newblock Johns Hopkins University Press, 4 edition, 2013.

\bibitem[Grimmett and Stirzaker(2001)]{Grimmett2001RandomProcesses}
Geoffrey Grimmett and David Stirzaker.
\newblock \emph{Probability and Random Processes}.
\newblock Oxford University Press, 3rd edition, 2001.

\bibitem[Harbrecht et~al.(2012)Harbrecht, Peters, and
  Schneider]{Harbrecht2012pivotedCholesky}
Helmut Harbrecht, Michael Peters, and Reinhold Schneider.
\newblock On the low-rank approximation by the pivoted {C}holesky
  decomposition.
\newblock \emph{Applied Numerical Mathematics}, 62\penalty0 (4):\penalty0
  428--440, 2012.

\bibitem[Hennig et~al.(2015)Hennig, Osborne, and Girolami]{HenOsbGirRSPA2015}
P.~Hennig, M.A. Osborne, and M.~Girolami.
\newblock Probabilistic numerics and uncertainty in computations.
\newblock \emph{Proceedings of the Royal Society of London A: Mathematical,
  Physical and Engineering Sciences}, 471\penalty0 (2179), 2015.

\bibitem[Liang et~al.(2015)Liang, Zou, Guo, Li, Zhang, Zhang, Huang, and
  Chen]{Liang2015pmDataset}
Xuan Liang, Tao Zou, Bin Guo, Shuo Li, Haozhe Zhang, Shuyi Zhang, Hui Huang,
  and Song~Xi Chen.
\newblock Assessing beijing's $pm_{2.5}$ pollution: severity, weather impact,
  apec and winter heating.
\newblock \emph{Proceedings of the Royal Society A: Mathematical, Physical and
  Engineering Sciences}, 471\penalty0 (2182):\penalty0 20150257, 2015.

\bibitem[Mnih(2008)]{Mnih2008EBSmasterThesis}
Volodymyr Mnih.
\newblock Efficient stopping rules.
\newblock Master's thesis, University of Alberta, Canada, 2008.

\bibitem[Mnih et~al.(2008)Mnih, Szepesv\'{a}ri, and
  Audibert]{Mnih2008EBstopping}
Volodymyr Mnih, Csaba Szepesv\'{a}ri, and Jean-Yves Audibert.
\newblock Empirical {B}ernstein stopping.
\newblock pages 672--679, 2008.

\bibitem[Mo{\v{c}}kus(1975)]{Mockus1975BO}
Jonas Mo{\v{c}}kus.
\newblock On {B}ayesian methods for seeking the extremum.
\newblock In Gury~I. Marchuk, editor, \emph{Optimization Techniques IFIP
  Technical Conference}, volume~27 of \emph{Lecture Notes in Computer Science},
  pages 400--404, 1975.

\bibitem[Moro et~al.(2014)Moro, Cortez, and Rita]{Moro2014BankMarketingDataset}
S{\'{e}}rgio Moro, Paulo Cortez, and Paulo Rita.
\newblock A data-driven approach to predict the success of bank telemarketing.
\newblock \emph{Decision Support Systems}, 62:\penalty0 22--31, 2014.

\bibitem[Popoviciu(1935)]{Popoviciu1935variance}
Tiberiu Popoviciu.
\newblock Sur les \'equations alg\'ebriques ayant toutes leurs racines
  r\'eelles.
\newblock \emph{Mathematica}, 9:\penalty0 129--145, 1935.

\bibitem[Rasmussen and Williams(2006)]{RasmussenWilliams}
Carl~E. Rasmussen and Christopher~K. Williams.
\newblock \emph{{Gaussian Processes for Machine Learning}}.
\newblock The {MIT} {P}ress, 2006.

\bibitem[Saibaba et~al.(2017)Saibaba, Alexanderian, and
  Ipsen]{Saibaba2017Determinant}
Arvind~K. Saibaba, Alen Alexanderian, and Ilse C.~F. Ipsen.
\newblock Randomized matrix-free trace and log-determinant estimators.
\newblock \emph{Numerische Mathematik}, 137\penalty0 (2):\penalty0 353--395,
  Oct 2017.

\bibitem[Seeger(2000)]{Seeger2000Skilling}
Matthias Seeger.
\newblock Skilling techniques for {B}ayesian analysis.
\newblock 2000.

\bibitem[Sharma et~al.(2010)Sharma, Gupta, and Kapoor]{Sharma2010variance}
Rajesh Sharma, Madhu Gupta, and Girish Kapoor.
\newblock Some better bounds on the variance with applications.
\newblock \emph{Journal of Mathematical Inequalities}, 4:\penalty0 355--363,
  2010.

\bibitem[Skilling(1989)]{Skilling1989Eigenvalues}
John Skilling.
\newblock \emph{The Eigenvalues of Mega-dimensional Matrices}, pages 455--466.
\newblock 1989.

\bibitem[Snelson and Ghahramani(2006)]{Snelson2006pseudo}
Edward Snelson and Zoubin Ghahramani.
\newblock Sparse {G}aussian processes using pseudo-inputs.
\newblock In Y.~Weiss, B.~Sch\"{o}lkopf, and J.~C. Platt, editors,
  \emph{Advances in Neural Information Processing Systems 18}, pages
  1257--1264. 2006.

\bibitem[Steinruecken et~al.(2019)Steinruecken, Smith, Janz, Lloyd, and
  Ghahramani]{Steinruecken2019automaticStatistician}
Christian Steinruecken, Emma Smith, David Janz, James Lloyd, and Zoubin
  Ghahramani.
\newblock The {A}utomatic {S}tatistician.
\newblock In Frank Hutter, Lars Kotthoff, and Joaquin Vanschoren, editors,
  \emph{Automated Machine Learning}, Series on Challenges in Machine Learning.
  2019.

\bibitem[Ubaru et~al.(2017)Ubaru, Chen, and Saad]{Ubaru2017fastTrA}
Shashanka Ubaru, Jie Chen, and Yousef Saad.
\newblock Fast estimation of $tr(f(a))$ via stochastic lanczos quadrature.
\newblock \emph{SIAM Journal on Matrix Analysis and Applications}, 38\penalty0
  (4):\penalty0 1075--1099, 2017.

\bibitem[{Wang} et~al.(2013){Wang}, {Zhang}, {Zhang}, and
  {Yi}]{Wang2013OpenBLAS}
Qian {Wang}, Xianyi {Zhang}, Yunquan {Zhang}, and Qing {Yi}.
\newblock {AUGEM}: {A}utomatically generate high performance {D}ense {L}inear
  {A}lgebra kernels on x86 {CPU}s.
\newblock In \emph{SC '13: Proceedings of the International Conference on High
  Performance Computing, Networking, Storage and Analysis}, pages 1--12, 2013.

\bibitem[Zhou(2002)]{Zhou2002coveringNumber}
Ding-Xuan Zhou.
\newblock The covering number in learning theory.
\newblock \emph{Journal of Complexity}, 18\penalty0 (3):\penalty0 739--767,
  2002.

\end{thebibliography}
\end{document}